\documentclass[aps,prd,a4paper,nosuperscriptaddress,nofootinbib,showpacs,twocolumn,showkeys,amsfonts,amssymb,amsmath]{revtex4-2}
\usepackage{amssymb,latexsym}

\usepackage{color}
\usepackage{amscd}
\usepackage{times}
\usepackage{epsfig}
\usepackage{psfrag}
\usepackage{graphicx}
\usepackage{amsthm,amsmath}
\usepackage{mathrsfs}
\usepackage{bm}
\usepackage[colorlinks=true, allcolors=blue]{hyperref}
\usepackage{tikz}
\usetikzlibrary{shapes.geometric, arrows.meta, decorations.pathmorphing}
\usepackage{leftindex}
\usepackage[dvipsnames]{xcolor}
\usepackage{physics}
\usepackage{pgfplots}
\usepackage{cleveref}
\pgfplotsset{compat=1.18} 

\begin{document}
\title[]{Witnessing entanglement between photon and matter due to graviton exchange }

\author{Arijit Dutta}
\affiliation{Center for Quantum Engineering, Research, and Education,
TCG CREST, Bidhan Nagar, Kolkata - 700091, India}

\author{Marko Toro\v{s} }
    \affiliation{Faculty of Mathematics and Physics, University of Ljubljana, Jadranska 19, SI-1000 Ljubljana, Slovenia}

\author{Sougato Bose}
    \affiliation{Department of Physics and Astronomy, University College London, London WC1E 6BT, United Kingdom}

\author{Anupam Mazumdar}
\affiliation{ 
Van Swinderen Institute, University of Groningen, 9747 AG Groningen, The Netherlands}

\begin{abstract}
The paper presents a scheme to detect entanglement arising from the quantum nature of gravity between a spin qubit and photons, using Stokes parameters. One of the crucial tests of the general theory of relativity is the bending of light due to the curvature. Recently, a quantum counterpart of this experiment to test the quantum nature of the gravitational interaction has been proposed, in which the spin-2, massless graviton yields entanglement between matter and a photon sector. Hence, it provides one of the most crucial experimental signatures for testing the quantum nature of gravity in a lab, since only spin-2-induced entanglement can yield the correct deflection of light due to matter. Here, we propose a positive partial-transpose (PPT) witness criterion for witnessing such an entanglement. We scan the entangled states in this context by studying the overlap of the final state, which is proportional to the entanglement  
phase. We exploit the Stokes observables to measure the photon state and the spins in the matter sector, thereby constructing a witness for the quantum nature of gravity in this setup. To quantify this entanglement, we will couple the photon to a local oscillator, whose phase need to be controlled to probe the orthogonal components of the macroscopic interference in the laser beam. We have shown that for a non-maximally entangled state mediated by the quantum nature of gravity, the witness attains a maximal negativity of $-0.052$. Our findings indicate that this witness effectively detects entanglement within the range
$0.71 \leq |\gamma| < 1$, where $\gamma$ is the overlap between the two coherent states of the photon, providing a clear signature of quantum correlations.
\end{abstract}

\maketitle

\section{Introduction}

Testing the quantum nature of gravity in a lab is one of the key outstanding challenges of modern physics, as it also tests the very foundations of quantum mechanics, e.g., to what extent macroscopic objects are quantum in nature~\cite{Bose:2025qns}. Now, there is a concrete proposal to test the quantum nature of gravitational interaction in a lab. We can now witness the entanglement between two matter sectors mediated via massless/massive spin-2 graviton, see the quantum gravity induced entanglement of matter (QGEM) protocol ~\cite{Bose:2017nin,ICTS,Elahi:2023ozf}, see also~\cite{Marletto:2017kzi}. For theoretical interpretation see~\cite{Marshman:2019sne,Bose:2022uxe,Vinckers:2023grv}. Historically, the light-bending due to curvature experiment was monumental in falsifying the theory of general relativity~\cite{Eddington}, because this phenomenon could not be explained by Newtonian gravity or by spin-0 mediation~\cite {Nordstrom:1913dga}. One requires the tensorial nature of the spin-2 gravitational interaction to explain the correct angle of light bending. Also, the fact that the virtual excitations of the graviton, which actually carry spin-2 and spin-0 degrees of freedom, differ from on-shell spin-2 with two helicity gravitational waves; see the discussion regarding the graviton propagator, how it behaves in general relativity, and various notable extensions \cite{Biswas:2011ar,Biswas:2013kla}.
For a quantised massless spin-2 graviton, its interaction with matter will lead to entanglement between matter and graviton~\cite{Rufo:2024ulr}, and similarly, a photon with graviton as shown in \cite{Biswas:2022qto}. However, to witness such an entanglement requires a scattering between matter and photon via a virtual graviton exchange,  as shown in this paper~\cite{Biswas:2022qto}, see also~\cite{Carney:2021vvt}.

Interestingly, the effective potential computed from scattering between the non-relativistic matter and the photon induced by the massless spin-2 graviton leads to the right deflection angle as that of general relativity, e.g. see~\cite{Scadron:2007qd}, and this very potential also leads to the entanglement between the position of the matter and the momentum of the photon mediated by the virtual graviton. 
In \cite{Biswas:2022qto}, we computed the entanglement entropy and sketched a scheme to detect such an entanglement as a possible test for the quantum nature of gravitational interaction with matter and photon.

In this paper, we will present the witnessing scheme in more detail by utilising the Stokes parameters, see~\cite{schnabel2003stokes,Agarwal2003Scheme,stokes1851composition},  on the optical side and the spin basis in the matter sector. Although, as we shall see, the experimental parameters are extremely demanding, our witnessing scheme now provides a way to test this entanglement in the lab, which was previously missing from the literature.

First, we describe the gravity-mediated interaction between the massive particle and the optical field. From this, we calculate the joint state of the combined system.
To detect gravity-induced entanglement, we derive an entanglement witness based on the Positive Partial Transpose (PPT) criterion~\cite{Peres:1996dw, horodecki1996separability}. For bipartite systems of dimensions $2\times2$ and $2\times3$, the PPT criterion provides a necessary and sufficient condition: a state is entangled if and only if the partial transpose of its density matrix yields at least one negative eigenvalue. In our case, identifying the eigenvector associated with the negative eigenvalue allows us to construct a projector whose partial transpose serves as the witness operator. If the expectation value of this operator is negative, the state is confirmed to be entangled. We then convert the witness into a set of experimentally measurable correlations. Specifically, we express the witness through correlations between the spin degrees of freedom embedded in the matter and the Stokes observables of the optical field.  The position state can be read via the spin system, for example, in the case of a defect embedded in the matter, such as the nitrogen vacancy (NV) centre in diamond, see~\cite{Doherty_2013}. We will need a material that is easy to read the spin on, either optically or magnetically. While the properties of the coherent state of the photon will be read by measuring the Stokes observables, see~\cite{stokes1851composition,Korolkova1996, korolkova2002polarization,zukowski}. There are schemes to create superposition with a NV-centered nanodiamonds, but the superposition sizes are typically very tiny in the: Stern-Gerlach type setup, of the order of $10-100 {\rm \mu m}$ for masses in the range ${\cal O}(10^{-14})$kg, see~\cite{Zhou:2022jug,Zhou:2022frl,Zhou:2022epb,Zhou:2024voj}. As we will see, we will need different techniques to create large spatial superpositions for massive objects in this scenario, see~\cite{Zhou:2024voj,Marshman:2018upe}. In this paper, we will not describe how to create a spatial superposition for much heavier masses, as we require here in this setup. We will keep them for a future discussion. 

Of course, decoherence will be a big challenge here~\cite{Hornberger:2008xkz,Arndt:2014blv,Romero-Isart:2011yun,Romero-Isart:2011sdw,Schut:2024lgp}. We will not delve too much on the decoherence due to collision, blackbody radiation and emission for the matter part. These have been addressed in \cite{Biswas:2022qto}. Furthermore, there will be dephasing due to vibrations, seismic noise~\cite{Toros:2020dbf}, electromagnetic noise~\cite{Schut:2023tce,Fragolino:2023agd}, and many other hosts of noise related to creating the matter-wave interferometer in the matter sector. 
Note that the coherence in the photon sector preserves the relative phase relationship, ensuring that the gravitationally induced phase difference remains measurable and does not average to zero due to dephasing~\cite{mandel1995optical, scully1997quantum}. However, stochastic fluctuations in the photon sector act as a source of decoherence~\cite{schlosshauer2019quantum, caves1981quantum}.

In Section II, we discuss the gravitational interaction between matter and photons. In Section III, we discuss the entanglement criterion and the parameters that lead to entanglement. In Section IV, we construct the witness for this entanglement.  In Section V, we conclude the paper.

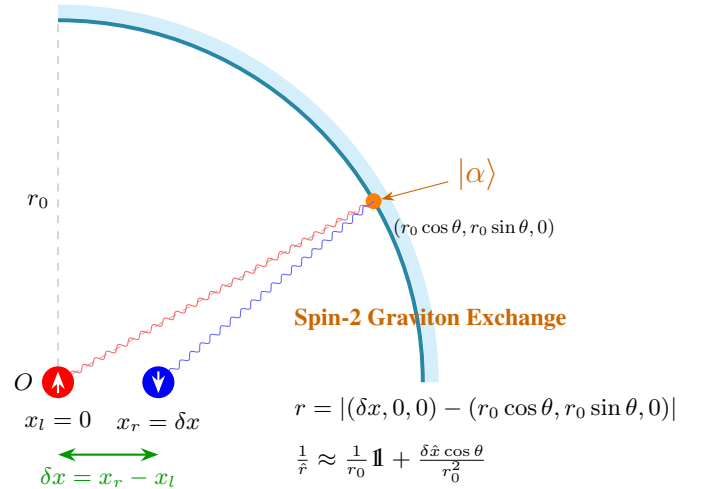
\begin{figure}[ht]
\centering
\begin{tikzpicture}[scale=1.2, font=\small, >=Stealth]

\coordinate (O) at (0,0);
\coordinate (L) at (0,0); 
\coordinate (R) at (1.1,0); 

\draw[line width=5pt, cyan!15] (0,4.1) arc (90:0:4.1); 
\draw[line width=1.5pt, cyan!60!black] (0,4) arc (90:0:4); 

\draw[dashed, gray!60] (O) -- (0,4) node[midway, left, black] {$r_0$};

\coordinate (P) at (3.46,2); 
\draw[dashed, gray!60] (O) -- (P);
\fill[orange] (P) circle (2.5pt);

\node[orange!80!black, font=\large] (alpha) at (4.6, 2.3) {$|\alpha\rangle$};
\draw[->, orange!80!black, shorten >=3pt] (alpha) -- (P);
\node[anchor=west, scale=0.8] at (3.6, 1.7) {$(r_0\cos\theta, r_0\sin\theta, 0)$};

\draw[decorate, decoration={snake, amplitude=1pt, segment length=5pt}, red!60] (L) -- (P);
\draw[decorate, decoration={snake, amplitude=1pt, segment length=5pt}, blue!60] (R) -- (P);

\fill[red] (L) circle (5pt);
\fill[blue] (R) circle (5pt);

\draw[->, white, thick] (L) ++(0,-0.12) -- ++(0,0.24);
\draw[<-, white, thick] (R) ++(0,-0.12) -- ++(0,0.24);

\node[left=6pt, black] at (O) {$O$};
\node[below=8pt, black] at (L) {$x_l = 0$};
\node[below=8pt, black] at (R) {$x_r = \delta x$};

\draw[<->, thick, green!60!black] (0,-0.8) -- (1.1,-0.8);
\node[green!60!black, below] at (0.55, -0.85) {$\delta x = x_r - x_l$};

\begin{scope}[shift={(2.5,-0.8)}]
    \node[anchor=west, align=left] at (0, 0.5) {
        $r = \left| (\delta x, 0, 0) - (r_0\cos\theta, r_0\sin\theta, 0) \right|$
    };
    \node[anchor=west, align=left] at (0, -0.1) {
        $\frac{1}{\hat r} \approx \frac{1}{r_0}\openone + \frac{\delta\hat{x}\cos\theta}{r_0^2}$
    };
    \node[orange!80!black, anchor=west] at (0, 1.5) {\textbf{Spin-2 Graviton Exchange}};
\end{scope}
\end{tikzpicture}
\caption{Schematic of the interaction between matter and photon. The left branch of the massive particle $x_l$ is positioned at the origin $O$. A Stern-Gerlach spin-dependent spatial superposition results in a spatial separation $\delta x$ ranging from ${\cal O}(1-100)\,\mu\mathrm{m}$, leading to a modified distance $\hat{r}$ to the optical field $|\alpha\rangle$, in a coherent state. By using Eq.~\eqref{strength}, and Eq.~\eqref{measure}, we show in Table~I~\ref{table}, for a coherent state with $|\alpha| = 10^{13}$ and an interaction time $\tau = 1\,\mathrm{s}$, the overlap $|\gamma|$ drops from $1.000$ to $0.6702$ for a $10\,\mathrm{kg}$ mass, signaling strong optomechanical entanglement witness between matter and photon due to a virtual massless spin-2 graviton exchange. In the presence of squeezing, one can maintain the same level of spatial separation $\delta x$ and distinguishability with a lower photon flux $|\alpha|^2$. }

\end{figure}

\section{Gravitational interaction between matter and photon}

In Ref.~\cite{Biswas:2022qto}, the authors propose a scheme to generate entanglement between a massive particle and an optical field mediated by the exchange of massless spin-2 virtual gravitons. The setup is an experiment protocol in which a massive particle of mass $m$ is harmonically trapped, while an optical field is confined to propagate along a half-ring trajectory of fixed radius $r$ around the massive particle. The massive particle is treated quantum mechanically and is initially prepared in a quantum ground state \(\ket{0}\) of the harmonic oscillator. The optical field is prepared in a strong coherent state \(\ket{\alpha}\). The gravitational interaction between the massive particle and the photons is modelled within a perturbative quantum field theoretic framework, see~\cite{Scadron:2007qd}, where gravity is mediated by virtual graviton exchange. Integrating out the gravitational field leads to an effective optomechanical interaction between the mechanical displacement of the massive particle and the photon number of the optical field. This interaction induces correlations between the massive particle and the photonic degrees of freedom and, under appropriate conditions, generates massive particle–photon entanglement. Below, we briefly discuss the scheme. 

At a quantum level, the gravitational interaction potential between the massive particle system and the optical field is given by~\footnote{Mathematically, the effective interaction potential $\hat V$ is obtained by taking the Fourier transform of the scattering amplitude associated with this combined process after integrating out the graviton propagator, see for the mathematical details of obtaining the quantum potential in \cite{Biswas:2022qto, Scadron:2007qd} .}
\begin{equation}
\hat{V} = -
\frac{2 G m (\hbar \omega)}{c^2~\hat r}
\,
\bm{\epsilon}^{*}_{\mathbf{k}',\nu'}
\cdot
\bm{\epsilon}_{\mathbf{k},\nu}
\,
\hat a^{\dagger}_{\mathbf{k}',\nu'}
\hat a_{\mathbf{k},\nu}.
\label{eq:quantumV}
\end{equation}
where $\omega$ is the optical frequency, $c$ is the velocity of light, and $\bm{\epsilon}_{\mathbf{k},\nu}$ is the polarization vector.
This interaction reproduces the gravitational light-bending effect in general relativity~\cite{Scadron:2007qd}.
The corresponding quantum operators are,
$\hat r,
\bm{\epsilon}_{\mathbf{k},\nu}\,\hat a_{\mathbf{k},\nu}$.
For the half-ring cavity configuration, the relative distance
between the photon and the harmonic oscillator are:
\begin{equation}
r
=
\left|
(\delta x,0,0)
-
(r_0\cos\theta, r_0\sin\theta,0)
\right|,
\end{equation}


where  $\delta x  \ll r_0$ represents the quantum fluctuation of the oscillator position. 
After expanding the inverse distance to first order in $\delta x,$ and integrating over the half-ring geometry, $\theta \in [-\pi/2,\pi/2]$, we obtain the operator-valued entities:
\begin{equation}
\frac{1}{\hat r}
\approx
\frac{1}{r_0}\openone
+
\frac{2\,\delta\hat x}{r_0^2}.
\end{equation}
Within this framework, we model the position fluctuation operator $\delta\hat{x}$ as an effective two-dimensional observable, spanned by two orthonormal eigenstates $\ket{l,\uparrow}$ and $\ket{r, \downarrow}$. For notational simplicity, we relabel these states as $\ket{l,\uparrow} \equiv \ket{l}$, and $\ket{r,\downarrow} \equiv \ket{r}$. These states correspond to two well-separated spatial wave packets localized around positions $x_l$ and $x_r$, respectively. In this reduced subspace, the operator $\delta\hat{x}$ admits
the spectral decomposition
\begin{equation}
\delta\hat{x}
=
x_l \ket{l}\!\bra{l}
+
x_r \ket{r}\!\bra{r},
\label{eq:deltaxlr}
\end{equation}
with
\begin{equation}
\label{eq:deltaxlr1}
\delta\hat{x}\ket{l} = x_l \ket{l},
\qquad
\delta\hat{x}\ket{r} = x_r \ket{r}.
\end{equation}
Substituting these results into the quantum light-matter interaction by virtue of massless graviton, yields the optomechanical interaction:
\begin{equation}
\hat V
\approx
-\frac{2Gm(\hbar \omega)}{c^2}
\left(
\frac{1}{r_0}\openone
+
\frac{2\,\delta\hat x}{r_0^2}
\right)
\bm{\epsilon}^{*}_{\mathbf{k}',\nu'}
\cdot
\bm{\epsilon}_{\mathbf{k},\nu}
\,
\hat a^{\dagger}_{\mathbf{k}',\nu'}
\hat a_{\mathbf{k},\nu}.
\end{equation}
The constant term produces an irrelevant global phase.
Retaining only the fluctuation-dependent contribution,
we obtain
\begin{equation}
\hat V
\simeq
-
\frac{4G m (\hbar \omega)}{c^2~r_0^{2}}
\,
\delta \hat x
\,
\bm{\epsilon}^{*}_{\mathbf{k}',\nu'}
\cdot
\bm{\epsilon}_{\mathbf{k},\nu}
\,
\hat a^{\dagger}_{\mathbf{k}',\nu'}
\hat a_{\mathbf{k},\nu}.
\label{eq:Vexpanded}
\end{equation}
To simplify analysis, we consider
$\mathbf{k}'=\mathbf{k},$ and $\nu'=\nu,$ such that for a single optical mode with fixed polarisation,
$\hat a^{\dagger}_{\mathbf{k},\nu}
\hat a_{\mathbf{k},\nu}
=\hat n.$ The interaction then reduces to the form
\begin{eqnarray}
\label{potential}
\hat V
=
-
\frac{4 G m (\hbar \omega)}{c^2~r_0^{2}}
\,
\delta \hat x
\,
\hat n.
\end{eqnarray}
Interestingly, from a field-theoretic perspective, such an interaction arises generically whenever a massive particle field couples to an optical field via graviton exchange. During the interaction, the stress-energy tensor of the massive particle acts as a source of the gravitational field, leading to the emission and absorption of a graviton in a scattering process.


\section{Entanglement criterion and parameters}

In our current scheme, we assume that the massive particle of mass $m$ is prepared in a spatial superposition of two distinct paths, labeled $l$ and $r$. Such a configuration can be realized by exploiting an internal spin degree of freedom of the massive particle. In particular, a Stern–Gerlach–type interferometric arrangement spatially separates the wave packet into two well-resolved trajectories that become correlated with different spin states, see~\cite{Marshman:2018upe,Marshman:2019sne, Zhou:2022epb,Zhou:2022frl}.
Such a scheme has been successfully implemented in an atom interferometer within the Stern-Gerlach setup, see~\cite{margalit2021realization,Folman2013,Folman2018,folman2019}.

This effective two-level description is justified when the spatial overlap between the left and right wave packets is negligible, such that the off-diagonal matrix elements $\bra{l}\delta\hat{x}\ket{r}$ can be safely neglected. The resulting potential is formally identical to the well-known cavity optomechanical interaction Hamiltonian~\cite{Biswas:2022qto}. Defining
\begin{equation}
\xi_s
=
\frac{4 G m (\hbar \omega) x_s}{c^2~r_0^{2}},
\end{equation}
the interaction Hamiltonian takes the form
\begin{eqnarray}
\label{Hamiltonian}
\hat H_{\text{int}}
=-\sum_{s}
\xi_s
\hat P_s
\hat n,
\end{eqnarray}
where $\hat P_s = \ket{s}\bra{s}$ denotes the projector onto the spatial branch $s$. We consider that initially, the combined (massive particle-photon) system is in a product state, 
\begin{eqnarray}
&|\psi(0)\rangle=\frac{1}{\sqrt{2}}\left(|l\rangle+|r\rangle\right)\otimes|\alpha\rangle.&
\label{tag703}
\end{eqnarray}
The system evolves under the Hamiltonian \(\hat{H}_{int}\) for an interaction time \(\tau\). The final state reads
\begin{eqnarray}
\label{state}
|\psi(\tau)\rangle&=&\text{exp}[-\frac{i}{\hbar}\,\hat{H}_{int}\,\tau]|\psi(0)\rangle.\nonumber\\
&=& \frac{1}{\sqrt{2}}
\left(
\ket{l}
\otimes
\ket{\alpha e^{i\phi_l}}
+
\ket{r}
\otimes
\ket{\alpha e^{i\phi_r}}
\right),
\end{eqnarray}
where \(\phi_s=\frac{\xi_s\tau}{\hbar}\). 
The coherent states $|\alpha e^{i\phi_l}\rangle$ and $|\alpha e^{i\phi_r}\rangle$ are
normalized but not orthogonal:
\begin{equation}
\label{overlap}
\gamma \equiv \langle\alpha e^{i\phi_l} |\alpha e^{i\phi_r} \rangle
=
\text{exp}[-|\alpha|^2(1-e^{i\Delta\phi})],
\qquad
0\leq|\gamma|\leq 1,
\end{equation}
where \(\Delta\phi=\phi_r-\phi_l \).
The overlap between the two coherent states is defined as $\gamma \equiv \langle\alpha e^{i\phi_l} |\alpha e^{i\phi_r} \rangle$:
\begin{eqnarray}
&\gamma = \exp\left[-|\alpha|^2(1-e^{i\Delta\phi})\right]& \nonumber \\
&= \exp\left[-|\alpha|^2(1 - \cos\Delta\phi - i \sin\Delta\phi)\right] &\nonumber \\
&= \exp\left[-|\alpha|^2(1 - \cos\Delta\phi)\right] e^{i |\alpha|^2 \sin\Delta\phi}&.
\label{gamma}
\end{eqnarray}
If $|\gamma|=1$, it means {\it no entanglement} between matter and photon. Hence, for the entanglement to be developed, we immediately notice that the $|\alpha|^2$ term has to be significantly large. Also, for a small $\Delta\phi,$ to ensure large entanglement between matter and photon due to massless spin-2 graviton exchange, we require:
\begin{equation}
\label{strength}
|\gamma|\approx \text{exp}[-|\alpha|^2\frac{\Delta\phi^2}{2}]
\end{equation}
The argument of the complex overlap, $\gamma$, in Eq.~(\ref{gamma})~it is given by:
\begin{eqnarray}
\label{phase}
\phi &= |\alpha|^2 \sin \Delta\phi
\end{eqnarray}
The gravitationally induced phase difference is given by:
\begin{equation}
\label{measure}
\Delta\phi = \frac{4 G m \omega (x_r-x_l)\tau}{c^2 r_0^2}.
\end{equation}
The linear entropy \(S_L\), which quantifies the mixedness  of the reduced state and serves as a measure of entanglement for pure bipartite states, is given by
\begin{equation}
\begin{aligned}
\label{entropy}
S_L &= 1 - \mathrm{Tr}(\rho_{\text{ph}}^2) \\
 &= \frac{1}{2}\left(1 - |\gamma|^2\right) \\
 &=\frac{1}{2}(1- \text{exp}[-|\alpha|^2 \Delta\phi^2]).
\end{aligned}
\end{equation}
Note that, from  Eq.~\eqref{entropy}, the linear entropy \(S_L \) is a monotonically decreasing function  of the  overlap \(|\gamma|\). 
Consequently, as the overlap \(|\gamma|\) increases, the bipartite entanglement decreases. 
In this effective bipartite qubit representation, the linear entropy $S_L$ ranges from $0$ for a separable state to a maximum of $1/2$. This upper bound corresponds to a maximally entangled state, where $|\gamma|=0$.

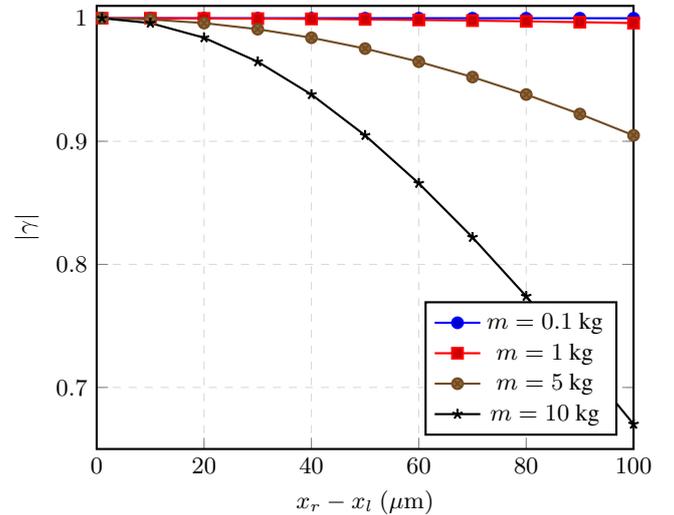
\begin{figure}[ht]
\centering
\begin{tikzpicture}
\begin{axis}[
    width=\linewidth,
    xlabel={$x_r - x_l~(\mu\mathrm{m})$},
    ylabel={$|\gamma|$},
    xmin=0, xmax=100,
    ymin=0.65, ymax=1.01,
    grid=major,
    grid style={dashed, gray!30},
    legend style={at={(0.97,0.03)},anchor=south east},
    thick,
]

\addplot coordinates {
(1,1.00000) (10,1.00000) (20,0.999998) (30,0.999996)
(40,0.999994) (50,0.999990) (60,0.999986)
(70,0.999980) (80,0.999974) (90,0.999968) (100,0.999960)
};
\addlegendentry{$m=0.1$ kg}

\addplot coordinates {
(1,1.00000) (10,0.999960) (20,0.99984) (30,0.99964)
(40,0.99936) (50,0.99900) (60,0.998561)
(70,0.998041) (80,0.997442) (90,0.996764) (100,0.996007)
};
\addlegendentry{$m=1$ kg}

\addplot coordinates {
(1,0.999990) (10,0.99900) (20,0.996007) (30,0.991037)
(40,0.984122) (50,0.975301) (60,0.964628)
(70,0.952164) (80,0.937983) (90,0.922166) (100,0.904804)
};
\addlegendentry{$m=5$ kg}

\addplot coordinates {
(1,0.999960) (10,0.996007) (20,0.984122) (30,0.964628)
(40,0.937983) (50,0.904804) (60,0.865842)
(70,0.821953) (80,0.774069) (90,0.723164) (100,0.670222)
};
\addlegendentry{$m=10$ kg}

\end{axis}
\end{tikzpicture}
\caption{Overlap $|\gamma|$ as a function of spatial separation $x_l-x_r$ for different masses (see Eq.~\eqref{strength}, and Eq.~\eqref{measure}). Parameters: $|\alpha|=10^{13}$, $\tau=1\,\mathrm{s}$, $r_0=0.25\,\mathrm{m}$, $\lambda=1\,\mu\mathrm{m}. $, we see how $|\gamma| $ scales with the superposition size. Note that $|\gamma|=1$ shows no entanglement; hence, for small values of $|\gamma|$, it is required to witness the entanglement, which prefers heavier masses to be kept in a large spatial superposition. }
\label{fig:gamma_plot}
\end{figure}

For the sake of illustration, we consider favourable parameters which can yield large entanglement, here
$G=6.674\times10^{-11}\,\mathrm{m^3.kg^{-1}.s^{-2}}$,
$\tau=1~\mathrm{s}$,
$r_0 = 0.25~\mathrm{m}$,
and $c = 3\times10^8~\mathrm{m/s}$.
For an optical wavelength $\lambda = 1~\mu\mathrm{m}$, the angular frequency is
\begin{equation}
\omega = \frac{2\pi c}{\lambda} \approx 1.88\times10^{15}~\mathrm{s^{-1}} .
\end{equation}
Numerical values of $|\gamma|$ are given in Table~\ref{table}, and the corresponding plot is shown in Fig.~\ref{fig:gamma_plot}~\footnote{To achieve detectable entanglement, one can tune the spatial separation $x_l - x_r$, mass $m$, and $r_0,$ to reduce the interaction time $\tau$ when employing strong lasers.}.

\begin{table*}[ht]
\centering
\label{table}
\begin{tabular}{c|ccccccccccc}
\hline
\hline
$m~(\mathrm{kg})$ 
& \multicolumn{11}{c}{$x_l - x_r~(\mu\mathrm{m})$} \\
\cline{2-12}
& 1 & 10 & 20 & 30 & 40 & 50 & 60 & 70 & 80 & 90 & 100 \\
\hline

0.1 
& 1.00000 
& 1.00000 
& 0.999998 
& 0.999996 
& 0.999994 
& 0.999990 
& 0.999986 
& 0.999980 
& 0.999974 
& 0.999968 
& 0.999960 \\

1 
& 1.00000 
& 0.999960 
& 0.99984 
& 0.99964 
& 0.99936 
& 0.99900
& 0.998561 
& 0.998041 
& 0.997442 
& 0.996764 
& 0.996007 \\

5 
& 0.999990 
& 0.999 
& 0.996007 
& 0.991037 
& 0.984122
& 0.975301
& 0.964628
& 0.952164 
& 0.937983
& 0.922166 
& 0.904804 \\

10 
& 0.999960 
& 0.996007 
& 0.984122
& 0.964628
& 0.937983
& 0.904804 
& 0.865842
& 0.821953
& 0.774069 
& 0.723164
& 0.670222 \\

\hline
\hline
\end{tabular}
\caption{Overlap $|\gamma|\approx \exp[-\tfrac{1}{2}|\alpha|^2 \Delta\phi^2]$ as a function of mass $m$ and spatial separation $x_l - x_r$ (see Eq.~\eqref{strength}, and Eq.~\eqref{measure}), showing the transition from weak to strong entanglement (see Eq.~\eqref{entropy}) for $|\alpha| = 10^{13}$ and $\tau = 1\,\mathrm{s}$. Other parameters are $r_0 = 0.25\,\mathrm{m}$ and $\lambda = 1\,\mu\mathrm{m}.$}
\end{table*}

\section{Constructing Entanglement Witness}
\label{mainfour}
To obtain a finite-dimensional matrix representation of the density
operator, we construct an orthonormal basis of the subspace
$\mathrm{span}\{|\alpha_1\rangle\equiv|\alpha e^{i\phi_l}\rangle, |\alpha_2\rangle\equiv|\alpha e^{i\phi_r}\rangle\}$ using the
Gram--Schmidt procedure~\cite{Gram1883,Schmidt1907,Leon2013}:
\begin{equation}
|e_1\rangle := |\alpha_1\rangle,
\qquad
|e_2\rangle :=
\frac{|\alpha_2\rangle - \gamma |\alpha_1\rangle}
{\sqrt{1-|\gamma|^2}}.
\end{equation}
These satisfy $\langle e_i|e_j\rangle=\delta_{ij}$, and
\begin{equation}
|\alpha_2\rangle = 
\gamma |e_1\rangle + \sqrt{1-|\gamma|^2}\,|e_2\rangle .
\end{equation}

Introducing the massive particle basis 
$|0\rangle\equiv|l\rangle$ and 
$|1\rangle\equiv|r\rangle$,
the total state \(|\psi(\tau)\rangle\)~\eqref{state} becomes
\begin{equation}
\label{ent}
|\Psi\rangle
=
\frac{1}{\sqrt2}
\left(
|0e_1\rangle
+
\gamma\,|1e_1\rangle
+
\sqrt{1-|\gamma|^2}\,|1e_2\rangle
\right),
\end{equation}
where $|se_j\rangle\equiv |s\rangle\otimes|e_j\rangle$.

The corresponding density operator $\rho=|\Psi\rangle\langle\Psi|$
has the following $4\times4$ matrix representation in the ordered basis
$\{|0e_1\rangle,|0e_2\rangle,|1e_1\rangle,|1e_2\rangle\}$:
\begin{equation}
\rho
=
\frac12
\begin{pmatrix}
\label{matrix}
1 & 0 & \gamma^* & \sqrt{1-|\gamma|^2} \\[4pt]
0 & 0 & 0 & 0 \\[4pt]
\gamma & 0 & |\gamma|^2 & \gamma \sqrt{1-|\gamma|^2} \\[4pt]
\sqrt{1-|\gamma|^2} & 0 &
\gamma^* \sqrt{1-|\gamma|^2} & 1-|\gamma|^2
\end{pmatrix}.
\end{equation}

In the orthonormal basis
${|0e_1\rangle, |0e_2\rangle, |1e_1\rangle, |1e_2\rangle}$, the partial transpose of the density matrix with respect to the massive particle subsystem is

\begin{equation}
\rho^{T_M}=
\frac12
\begin{pmatrix}
\label{matrixt}
1 & 0 & \gamma & 0 \\
0 & 0 & \sqrt{1-|\gamma|^2} & 0 \\
\gamma^* & \sqrt{1-|\gamma|^2} & |\gamma|^2 &
\gamma\sqrt{1-|\gamma|^2} \\
0 & 0 &
\gamma^*\sqrt{1-|\gamma|^2} & 1-|\gamma|^2
\end{pmatrix}.
\end{equation}
The partial transpose \(\rho^{T_M}\) possesses a single negative eigenvalue.
Explicitly, one finds
\begin{equation}
\lambda_{\text{neg}}
=
-\frac{1}{2}\sqrt{1-|\gamma|^{2}},
\end{equation}
which is strictly negative whenever
$|\gamma| < 1$. This establishes the presence of entanglement via the Peres--Horodecki criterion, \cite{Peres:1996dw,horodecki1996separability}.
However, although the negativity provides a formal entanglement witness, it is not directly accessible experimentally. In the following, we therefore propose an operational method to detect entanglement in the state \eqref{ent} by measuring the spin of the massive particle and the Stokes observables of the photons.

We express the complex overlap as
\(
\gamma = |\gamma|e^{i\phi}\),
where \(0 \le |\gamma| \le 1\).
The rank-one projector associated with the negative eigenvalue of the
partially transposed density matrix is
\begin{equation}
\Pi_- = \frac{1}{2(1+\alpha'^2)}
\begin{pmatrix}
1
& \alpha' e^{i\phi}
& -\alpha' e^{i\phi}
& e^{2i\phi}
\\[4pt]
\alpha' e^{-i\phi}
& \alpha'^2
& -\alpha'^2
& \alpha' e^{i\phi}
\\[4pt]
-\alpha' e^{-i\phi}
& -\alpha'^2
& \alpha'^2
& -\alpha' e^{i\phi}
\\[4pt]
e^{-2i\phi}
& \alpha' e^{-i\phi}
& -\alpha' e^{-i\phi}
& 1
\end{pmatrix},
\end{equation}
where the real parameter
\begin{eqnarray}
\label{alphaprime}
  \alpha' = \frac{1+\sqrt{1-|\gamma|^2}}{|\gamma|}.
\end{eqnarray}

Taking the partial transpose with respect to the massive particle subsystem $M$, in the
 basis
$\{|0e_1\rangle,|0e_2\rangle,|1e_1\rangle,|1e_2\rangle\}$, we obtain
\begin{equation}
\Pi_-^{T_M}
=
\frac{1}{2(1+\alpha'^2)}
\begin{pmatrix}
1
& \alpha' e^{i\phi}
& -\alpha' e^{-i\phi}
& -\alpha'^2
\\[4pt]
\alpha' e^{-i\phi}
& \alpha'^2
& e^{-2i\phi}
& \alpha' e^{-i\phi}
\\[4pt]
-\alpha' e^{i\phi}
& e^{2i\phi}
& \alpha'^2
& -\alpha' e^{i\phi}
\\[4pt]
-\alpha'^2
& \alpha' e^{i\phi}
& -\alpha' e^{-i\phi}
& 1
\end{pmatrix}.
\end{equation}


To derive an observable witness, we define the effective photon-qubit Pauli operators in the orthonormal basis~$\{|e_1\rangle,|e_2\rangle\}$ as
\begin{eqnarray}
\label{basis}
\sigma_0^{(e)} &=& |e_1\rangle\langle e_1| + |e_2\rangle\langle e_2|\nonumber\\
\sigma_x^{(e)} &=& |e_1\rangle\langle e_2| + |e_2\rangle\langle e_1|\nonumber\\
\sigma_y^{(e)}&=& -i\,|e_1\rangle\langle e_2| + i\,|e_2\rangle\langle e_1|\nonumber\\
\sigma_z^{(e)} &=& |e_1\rangle\langle e_1| - |e_2\rangle\langle e_2|.
\end{eqnarray}
The entanglement witness \(\Pi_-^{T_M}\) can then be expanded as: 
\begin{widetext}
\begin{eqnarray}
\label{witness}
\Pi_-^{T_M}
&=\frac{1}{4(1+\alpha'^2)}
\Big[
(1+\alpha'^2)\,\sigma_0\otimes\sigma^{(e)}_0
+
(1-\alpha'^2)\,\sigma_z\otimes\sigma_z^{(e)}
+
(\cos(2\phi)-\alpha'^2)\,\sigma_x\otimes\sigma_x^{(e)}
+
(\cos(2\phi)+\alpha'^2)\,\sigma_y\otimes\sigma_y^{(e)}\nonumber\\
&-\sin(2\phi)
\big(
\sigma_x\otimes\sigma_y^{(e)}
-
\sigma_y\otimes\sigma_x^{(e)}
\big)
+2\alpha'\cos\phi
\big(
\sigma_z\otimes\sigma_x^{(e)}
-
\sigma_x\otimes\sigma_z^{(e)}
\big)
-2\alpha'\sin\phi
(\sigma_y\otimes\sigma_z^{(e)}
+
\sigma_z\otimes\sigma_y^{(e)})
\Big].
\end{eqnarray}
\end{widetext}
The phase is determined by
\(
\phi = \arg(\gamma).
\)
Using the real and imaginary parts
\(
\gamma= \Re(\gamma) + i\,\Im(\gamma),\)
we have
\(
\cos\phi= \frac{\Re(\gamma)}{|\gamma|}\),
and 
\(\sin\phi = \frac{\Im(\gamma)}{|\gamma|}.
\)

While the expansion in Eq.~\eqref{witness} is expressed in terms of the abstract effective Pauli operators $\sigma^{(e)}_i$, these are not directly accessible for a single-mode field. To obtain an observable witness, we exploit the fact that in the single-mode regime, the normalized Stokes operators~\cite{zukowski} (see Appendix~\ref{appB} for details) $\hat{S}'_i$ act as the physical counterparts of the logical qubit operators at the level of expectation values. By identifying the mapping 
\begin{eqnarray}    
\langle\sigma^{(e)}_z \rangle \to\langle\hat{S}'_1\rangle,~
\langle\sigma^{(e)}_x \rangle \to \langle\hat{S}'_2\rangle,~
\langle\sigma^{(e)}_y\rangle \to \langle\hat{S}'_3\rangle,
\end{eqnarray}
the measured entanglement witness~\eqref{witness} value can be expressed in the follwoing concise form:
\begin{equation}
\label{Tmeas}
\langle \Pi^{T_M}_- \rangle_{meas} = \frac{1}{4(1 + \alpha'^2)} \left[ \mathcal{T}_0 + \mathcal{T}_{S1} + \mathcal{T}_{S2} + \mathcal{T}_{S3} \right],
\end{equation}
where the individual contributions from each sector are:
\begin{itemize}
    \item $\mathcal{T}_0 = (1 + \alpha'^2) \langle \sigma_0 \otimes \hat{S}'_0 \rangle$
    \item $\mathcal{T}_{S_1}=(1 - \alpha'^2) \langle \sigma_z \otimes \hat{S}'_1 \rangle - 2\alpha' \left( \cos \phi \langle \sigma_x \otimes \hat{S}'_1 \rangle + \sin \phi \langle \sigma_y \otimes \hat{S}'_1 \rangle \right)$
    \item $\mathcal{T}_{S_2} = (\cos 2\phi - \alpha'^2) \langle \sigma_x \otimes \hat{S}'_2 \rangle + \sin 2\phi \langle \sigma_y \otimes \hat{S}'_2 \rangle$
    \item $\mathcal{T}_{S_3} = (\cos 2\phi + \alpha'^2) \langle \sigma_y \otimes \hat{S}'_3 \rangle - \sin 2\phi \langle \sigma_x \otimes \hat{S}'_3 \rangle - 2\alpha' \sin \phi \langle \sigma_z \otimes \hat{S}'_3 \rangle$
\end{itemize}
All correlation functions are explicitely calculated in Appendix~\ref{appC}~(see Eqs.~\eqref{calcor}, \eqref{calcorF}).

{\it \underline{Separability}}: To verify the witness threshold, we consider a separable state where the matter qubit and optical sector are uncoupled, i.e., $\Delta \phi = 0$~(see Eq.~\eqref{overlap}). Given that the witness \eqref{Tmeas} is state-dependent, we evaluate it using the parameters for a separable state: the phase shift is uniform $\phi_l = \phi_r = \phi_m$~(see Eq.~\eqref{state}), overlap magnitude $|\gamma|=1$ ~(see Eq.~\eqref{strength}), phase angle $\phi = 0$~(see Eq.~\eqref{phase}), and consequently~$\alpha' = 1$~(see Eq.~\eqref{alphaprime}). 
Under these conditions, non-zero contributions are obtained from only two sectors of the measured witness, specifically $\mathcal{T}_0=2, \mathcal{T}_1=-2.$ Hence, the expectation value of the witness~\eqref{Tmeas} for the separable state is exactly {\it zero}.

In Appendix~\ref{appD}, the non-negativity of the witness is demonstrated for a general separable state within our interaction model, confirming that the entanglement threshold is never breached in the absence of quantum correlations.

{\it\underline{Negativity of the witness}}:
We analyze a matter-photon system where the interaction generates a non-maximally entangled state, where \(0<|\gamma|<1\). In Appendix~\ref{appC}, it is shown that (see Eqs.~\eqref{calcor}, and \eqref{calcorF}) for strong optical fields~(\(|\alpha|\gg 1)\), and small phase differences~\(\Delta\phi\), the optimization of Local oscillator (LO) phase $\phi_\beta$~\footnote{In this high-intensity balanced homodyne configuration, the macroscopic coherent state $|\beta \rangle$, where~\(\beta=|\beta|e^{i\phi_\beta}\),~ serves as a stable classical phase reference. The large photon flux ensures that the reference phase $\phi_\beta$ can be precisely tuned to rotate the reference frame of the macroscopic interference. This allows the physical photon-number difference at the output ports to selectively probe the real and imaginary components of the phase-sensitive coherence of the input signal.}, reduces the measured witness~\eqref{Tmeas} to {\it four} surviving correlation functions. Specifically, we obtain:
\begin{itemize}
    \item $\mathcal{T}_0 = (1 + \alpha'^2)$
    
    \item $\mathcal{T}_{S_1}
    = -2|\gamma|\alpha' \left( \cos^2\phi + \sin^2\phi \right)
    = -2|\gamma|\alpha'$
    
\item 
$\mathcal{T}_{S_3}
= -|\gamma|(\cos 2\phi + \alpha'^2)\sin\phi
   + |\gamma|\sin 2\phi \cos\phi\nonumber \\
= |\gamma|(1 - \alpha'^2)\sin\phi $.
\end{itemize}
Summing all terms yields: 
\begin{equation}
\label{finalW}
\langle \Pi^{T_M}_- \rangle_{meas}
= \frac{1}{4(1 + \alpha'^2)}
\Big[
(1 + \alpha'^2)
- 2|\gamma|\alpha'
+ |\gamma|(1 - \alpha'^2)\sin\phi
\Big].
\end{equation}
This expression explicitly demonstrates that for non-maximally entangled states, the witness remains phase-dependent. 
The intrinsic phase 
\(\phi=\text{arg}(\gamma)\) is fixed by the interaction and is not directly tunable. However, the experimentally measured normalized Stokes observables depend on the chosen LO phase~\(\phi_\beta\). By scanning~\(\phi_\beta\), one can identify the extremal values of the phase-sensitive correlation functions, which operationally minimizes the witness~\eqref{finalW}.

For example, We evaluate the expectation value of the witness for the specific parameter \(|\gamma|=0.9,\) and the corresponding \(\alpha'~(1.595)\)~(see Eq.~\eqref{alphaprime}). 
Substituting these values into the general expression~\eqref{finalW}, we get:
\begin{equation}
\label{finalW0}
\langle \Pi^{T_M}_- \rangle_{meas}
\approx 0.25-0.203-0.098\,\sin\phi.
\end{equation}

By sweeping the LO phase, this correlator exhibits a sinusoidal oscillation, allowing for the identification of its maximum magnitude. When this correlator is maximized, e.g.\ $\langle \sigma_y \otimes \hat S'_3 \rangle \approx -0.9$ for $|\gamma|=0.9$, the condition $\sin\phi \to 1$ is effectively achieved through measurement optimization rather than direct control of the intrinsic phase $\phi$.
Substituting this optimized peak correlator into Eq.~\eqref{finalW0} results in a final witness value of
\(\langle \Pi^{T_M}_- \rangle_{meas}=-0.051\).

Next we plot the measured witness value~\eqref{finalW} as a function of the overlap magnitude \(|\gamma|\)~(see Eqs.~\eqref{overlap}, and \eqref{strength}). 
\begin{figure}[h]
\centering
\begin{tikzpicture}
\label{figW}
\begin{axis}[
    width=9cm,
    height=7cm,
    xlabel={$|\gamma|$},
    ylabel={$\langle \Pi^{T_M}_- \rangle_{\mathrm{meas}}$},
    grid=major,
    xmin=0.7, xmax=1,
    ymin=-0.06, ymax=0.0,
    thick,
    mark size=2.5pt,
]

\addplot[
    mark=*,
]
coordinates {
      (1,0)
     (0.99,-0.03)
    (0.95,-0.05)
    (0.924,-0.052)
    (0.90,-0.051)
    (0.85,-0.043)
    (0.80,-0.03)
    (0.75,-0.015)
    (0.71,-0.001)
    (0.7,0)
};

\addplot[
    only marks,
    mark=*,
    mark size=3.5pt,
]
coordinates {
    (0.924,-0.052)
};

\end{axis}
\end{tikzpicture}
\caption{
Entanglement witness $\langle \Pi^{T_M}_- \rangle_{\mathrm{meas}}$ as a function of the coherent-state overlap $|\gamma|$. 
The witness exhibits a non-monotonic dependence, attaining maximal negativity  
around $|\gamma| \approx 0.92$, reflecting a trade-off between distinguishability 
and measurement sensitivity.
}
\label{fig:witness_gamma}

\end{figure}
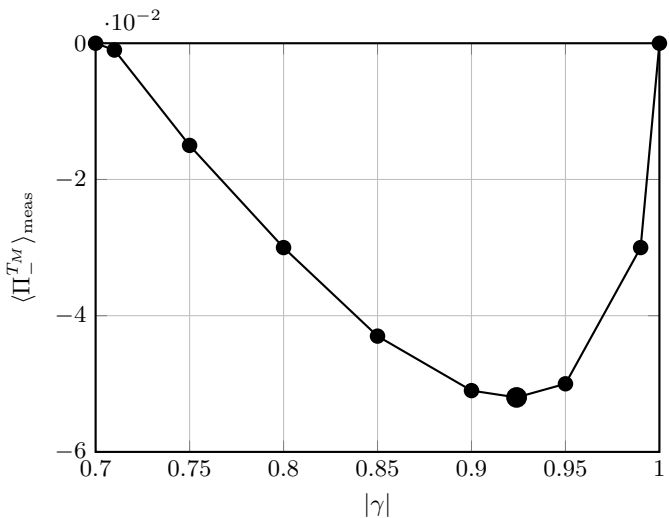
As illustrated in Fig.~\ref{fig:witness_gamma},~the witness effectively detects entanglement in the matter-photon state~\eqref{state} for the regime $0.71\leq|\gamma|<1$. The witness reaches a maximum negativity of $-0.052$ at an overlap magnitude $|\gamma|=0.924.$

Note that this specific witness is not efficient for detecting entanglement in states with low overlap values  \(|\gamma|\to 0\). However, in the regime of gravity-induced interactions, the coupling between the matter and the optical field is inherently weak, and the resulting phase difference \(\Delta\phi\)~(see Eqs.~\eqref{state}, \eqref{phase}, and \eqref{measure})~between two optical branches is exceedingly small. Consequently, the entanglement generated remains in the weak regime where~\(|\gamma|\) is close to unity~(see Table~I~\ref{table}), precisely where this interferometric witness exhibits its highest sensitivity.

\section{Conclusion}
In summary, we have proposed an experimental protocol to detect gravity-mediated entanglement between a massive particle and an optical field in a coherent state. By employing a Stern-Gerlach-type configuration, we generate a spatial superposition of the particle's paths correlated with its spin states. Simultaneously, by utilising normalized Stokes operators, our protocol ensures that the continuous outcome observables of the high-intensity field are mapped onto a bounded interval between $-1$ and $+1$ at the level of expectation values. Our analysis reveals that the strength of this entanglement is governed by the overlap parameter $|\gamma|$. Due to the extreme weakness of the gravitational constant ($G/c^2$), achieving measurable correlations requires a high-intensity laser pulse with approximately $10^{13}$ photons. 

As demonstrated in Table~I~\ref{table}, for a $10\,\mathrm{kg}$ mass and a $1\,\mathrm{s}$ interaction time, the system undergoes a clear transition from weak to strong entanglement as the spatial separation increases; the overlap $|\gamma|$ drops from $0.9999$ at $1\,\mu\mathrm{m}$ to $0.6702$ at $100\,\mu\mathrm{m}$. To quantify this entanglement, we have derived a potent PPT witness constructed from hybrid spin-Stokes observables. By optimizing the "Local Oscillator" phase to probe the orthogonal components of the macroscopic interference, we have shown that for a non-maximally entangled state with $|\gamma|=0.924$, the witness attains a maximal negativity of $-0.052$. Our findings indicate that this witness effectively detects entanglement within the range $0.71 \leq |\gamma| < 1$, providing a clear signature of quantum correlations. Furthermore, in Appendix~\ref{appE}, we extend our analysis to include the effects of isotropic noise, identifying a critical noise parameter 
\(v_{\text{critical}} = \frac{1}{1 + 2\sqrt{1 - |\gamma|^2}}\),~
above which entanglement is preserved. This analysis demonstrates that maximizing subsystem distinguishability provides the most robust protection against decoherence, with entanglement persisting to the absolute limit of $v > 1/3$. There will be other sources of decoherence besides the isotropic noise, fluctuations in the photon sector, spin-read out and other sources of dephasing, which we still need to analyse. However, we  now have a framework to discuss the entanglement witness by including all these sources of noise in the analysis.

Also, implementing this scheme presents significant experimental challenges, which we will have to work out meticulously. Beyond the daunting task of maintaining a $100\,\mu\mathrm{m}$ spatial superposition for a $10\,\mathrm{kg}$ mass over a $1\,\mathrm{s}$ duration, the detection of photons to measure Stokes observables is equally challenging for such a high-power laser. Precise measurement of these observables necessitates detectors capable of handling an immense photon flux without saturation while maintaining the sensitivity required to extract the subtle gravitational signature. Despite these obstacles, this hybrid matter-light interface provides a concrete pathway to observing quantum gravity effects in a laboratory setting. The spin-Stokes witness developed here serves as a versatile tool for identifying entanglement in macroscopic systems, offering a robust framework that can be extended to various other experimental protocols.

\begin{acknowledgments}
SB would like to acknowledge EPSRC grants
(EP/N031105/1, EP/S000267/1, and EP/X009467/1)
and grant ST/W006227/1.
 S.B and A.M.'s research are funded by the Gordon and Betty Moore Foundation through Grant GBMF12328, DOI 10.37807/GBMF12328. This material is based on work supported by the Alfred P. Sloan Foundation under Grant No. G-2023-21130
\end{acknowledgments}

\appendix
\section{Stokes Operators, Effective Optical Qubit, and Experimental Implementation}
\label{appA}
In 1851, Stokes~\cite{stokes1851composition} introduced four parameters to quantify the polarization state of classical light. Since then, various generalizations have appeared in the literature~\cite{Wolf1959, Jauch1980, Korolkova1996, BornWolf1999, Abouraddy2002, Agarwal2003Scheme}.  
In this Appendix, we summarize the interferometric realization of the quantum version of the Stokes operators~\cite{korolkova2002polarization, schnabel2003stokes, dutta2021robust}, their photon-number interpretation, and their role in implementing the effective optical qubit appearing in the entanglement witness.

{\bf Interferometric Definitions}:
 For the measurement of the optical sector, we employ a balanced homodyne configuration operated in the strong-field regime. Unlike standard homodyne detection designed for weak signals (see Refs.~\cite{leonhardt_measuring_1997, welsch1999homodyne} for a review), this setup allows for the direct reconstruction of the normalized Stokes operators by interfering two bright coherent states~\cite{korolkova2002polarization, schnabel2003stokes}. Operating with comparable intensities in both branches allows for inherently high interference visibility, ensuring that the phase-sensitive information required for the entanglement witness is extracted with maximal contrast. By mapping the relative phase of these macroscopic fields onto the intensity difference of the output modes, we can precisely extract the bipartite correlations necessary to satisfy the witness condition.

In this framework, a balanced 50:50 beam splitter mixes a signal mode $\hat{a}$ and a local oscillator (LO) mode $\hat{b}$, with a relative phase $\theta_{LO}$. The output modes of the beam splitter, $\hat{a}_1$ and $\hat{a}_2$, are defined via the transformation:
\begin{align}
\hat{a}_1 &= \frac{1}{\sqrt{2}} \left( \hat{b} + e^{i\theta_{LO}}\hat{a} \right) \nonumber \\
\hat{a}_2 &= \frac{1}{\sqrt{2}} \left( \hat{b} - e^{i\theta_{LO}}\hat{a} \right) \label{A1}
\end{align}
The physical mechanism enabling this mapping is the coherent interference of the macroscopic signal and local oscillator fields. This process converts the relative phase of the input modes—which carries the signature of the matter-photon interaction—into a measurable differential photon flux at the output detectors. The visibility $V$ quantifies the contrast of this interference; in the regime of comparable intensities, $V \approx 1$ ensures that the phase-sensitive correlations required for the witness are extracted with maximal contrast.

The Stokes observables are defined in the two bosonic output modes $(\hat{a}_1, \hat{a}_2)$ as~\cite{WallsMilburn2008}:
\begin{align}
\hat{S}_0 &= \hat{a}_1^\dagger \hat{a}_1 + \hat{a}_2^\dagger \hat{a}_2 \nonumber \\
\hat{S}_1 &= \hat{a}_1^\dagger \hat{a}_1 - \hat{a}_2^\dagger \hat{a}_2 \nonumber \\
\hat{S}_2 &= \hat{a}_1^\dagger \hat{a}_2 + \hat{a}_2^\dagger \hat{a}_1 \nonumber \\
\hat{S}_3 &= i(\hat{a}_2^\dagger \hat{a}_1 - \hat{a}_1^\dagger \hat{a}_2) \label{A2}
\end{align}

By substituting the transformation in Eq.~\eqref{A1} into Eq.~\eqref{A2}, the Stokes observables can be expressed in terms of the input signal and LO modes:
\begin{align}
\hat{S}_0 &= \hat{b}^\dagger\hat{b} + \hat{a}^\dagger\hat{a} \nonumber \\
\hat{S}_1(\theta_{LO}) &= \hat{b}^\dagger\hat{a} e^{i\theta_{LO}} + \hat{a}^\dagger\hat{b} e^{-i\theta_{LO}} \nonumber \\
\hat{S}_2 &= \hat{b}^\dagger\hat{b} - \hat{a}^\dagger\hat{a} \nonumber \\
\hat{S}_3(\theta_{LO}) &= i(\hat{b}^\dagger\hat{a} e^{i\theta_{LO}} - \hat{a}^\dagger\hat{b} e^{-i\theta_{LO}})
\label{A3}
\end{align}

Note that while $\hat{S}_1$ represents the population difference in the output mode basis $(\hat{a}_1, \hat{a}_2)$, the beam splitter transformation maps this observable onto the phase-sensitive coherence of the input signal and local oscillator modes $(\hat{a}, \hat{b})$. In this framework, the strong coherent state of the local oscillator acts as a stable, non-entangled phase reference for the photon sector, providing a stable phase-space reference to map its continuous-variable signal quadratures onto an effective discrete-variable (qubit) basis.

\section*{Phase Control for Stokes Observables}

In the experimental setup, the primary measured quantity is the photon-number difference at the output of the beam splitter, which corresponds to $\hat{S}_1$~(see Eq.~\eqref{A2}). By varying $\theta_{LO}$ within $\hat{S}_1(\theta_{LO})$~(see Eq.~\eqref{A3}), we can access {\it{two}} Stokes observables:
\begin{enumerate}
    \item To measure the interference component corresponding to $\hat{S}_1$ (at a reference phase), we set $\theta_{LO} = 0$, giving $\hat{S}_1(0) = \hat{b}^\dagger\hat{a} + \hat{a}^\dagger\hat{b}$.
    \item To measure the component corresponding to $\hat{S}_3$, we set $\theta_{LO} = \pi/2$, which transforms the interference term:
    \begin{equation}
    \hat{S}_1(\pi/2) = i(\hat{b}^\dagger\hat{a} - \hat{a}^\dagger\hat{b}) = \hat{S}_3(0). \tag{A4}
    \end{equation}
    \end{enumerate}
The operator $\hat{S}_2$ represents the number difference between the LO and signal modes themselves ($\hat{b}^\dagger\hat{b} - \hat{a}^\dagger\hat{a}$) and is measured by bypassing the beam splitter.
These operators satisfy the $SU(2)$ commutation relations $[\hat{S}_i, \hat{S}_j] = 2i\epsilon_{ijk} \hat{S}_k,$ where $i,j,k \in \{1,2,3\}$. Thus, all Stokes observables associated with the input modes of the beam splitter are expressed as measurable photon-counting observables at the output modes.

\section{Effective Optical Qubit Mapping and Experimental Estimation}
\label{appB}
To estimate the entanglement witness, we use a mapping between the normalized Stokes operators and the effective Pauli operators of the qubit subspace at the level of expectation values (see Sec.~\ref{mainfour} in main text). The use of normalized Stokes operators ensures that the measured observables remain bounded and encode only relative phase information, enabling a consistent qubit mapping. In the limit of high total photon numbers, we utilize the ratio of expectations to define these normalized observables:
\begin{equation}
\langle \hat S'_j \rangle = \frac{\langle \hat S_j \rangle}{\langle \hat S_0 \rangle},
\label{B1}
\end{equation}
where $\hat S_j$ and $\hat S_0$ represent the photon number difference and total photon count, respectively. 

In this high-photon regime ($|\alpha|, |\beta| \gg 1$), the probability of measuring zero total photons is exponentially suppressed ($\sim e^{-\langle n \rangle}$), ensuring the ratio is always well-defined. Furthermore, since $|\hat S_j| \le \hat S_0$ by definition, the resulting expectations are strictly bounded within the interval $[-1, 1]$. This boundedness allows the optical field's phase space to be mapped onto a legitimate qubit basis without violating the algebraic constraints of the witness. 


The joint matter-photon correlation functions are reconstructed by normalizing the accumulated bipartite products by the total photon flux recorded over $N$ trials:
\begin{equation}
\langle \sigma_i \otimes \hat S'_j \rangle = \frac{\sum_{k=1}^N m_i^{(k)} S_j^{(k)}(\theta_{LO})}{\sum_{k=1}^N S_0^{(k)}}.
\label{B2}
\end{equation}
Here, $m_i^{(k)}$ is the matter qubit measurement outcome and $S_j^{(k)}$ is the unnormalized Stokes value for the $k$-th run. By measuring these operators through homodyne detection against a Local Oscillator, we extract the bipartite correlations necessary to satisfy the witness condition. This representation allows the complex gravitational interaction—encoded in the physical phase $\Delta\phi$ and the overlap $\gamma$—to be treated within the standard framework of quantum information theory.

\section{Rigorous Implementation via Normalized Stokes Operators} \label{appC}

Consider the combined state of the bipartite system and the Local Oscillator (LO):
\begin{equation} 
|\Psi\rangle = \frac{1}{\sqrt{2}} \left( |0\rangle |\alpha_1, \beta\rangle + |1\rangle |\alpha_2, \beta\rangle \right), 
\end{equation}
where the signal and LO complex amplitudes are defined as:
\begin{equation*} 
\alpha_1 = |\alpha| e^{i\phi_l}, \quad \alpha_2 = |\alpha| e^{i\phi_r}, \quad \text{and} \quad \beta = |\beta| e^{i\phi_\beta}.
\end{equation*}
In the semi-classical limit, both the signal and the local oscillator contain large photon numbers ($|\alpha|^2, |\beta|^2 \gg 1$). 

 The correlation functions involving the qubit operator $\sigma_x$ depend on matrix elements of the form:
\begin{equation} 
\langle 0, \alpha_1, \beta | \sigma_x \otimes \hat{S}'_j | 1, \alpha_2, \beta \rangle = \langle \alpha_1, \beta | \frac{\hat{S}_j}{\hat{S}_0} | \alpha_2, \beta \rangle. \label{C2} 
\end{equation}
In this high-intensity regime, the expectation value of the joint observable is represented in the following form:

\begin{equation} 
\langle \sigma_x \otimes \hat{S}'_j \rangle = \frac{\langle \alpha_1, \beta | \hat{S}_j | \alpha_2, \beta \rangle + \text{h.c.}}{2 \langle \Psi | \hat{S}_0 | \Psi \rangle}, \label{C3}  
\end{equation}
where the denominator $\langle \Psi | \hat{S}_0 | \Psi \rangle=|\alpha|^2 + |\beta|^2$ corresponds to the total intensity. This framework allows for the extraction of bipartite correlations by measuring the intensity differences at the output of the beam splitter.

All correlation functions are evaluated on identically prepared states. In the following, $V=\frac{2|\alpha||\beta|}{|\alpha|^2+|\beta|^2}$ represents the visibility and $\bar{\phi}$ is the average interaction phase, defined as $\bar{\phi}=\frac{1}{2}(\phi_l+\phi_r)$. The total controllable phase is $\Theta = \theta_{LO} + \phi_\beta$. Here, $\phi_\beta^{(1)}$ is chosen as the LO phase to estimate the expectation values associated with $\hat{S}'_1$, while $\phi_\beta^{(3)}$ is used for $\hat{S}'_3$. The correlation functions for the joint matter-photon state—evaluated at the reference phase $\theta_{LO}=0$ as defined in Eq.~\eqref{A3}—are given as follows:
\begin{eqnarray} 
\label{calcor}
\langle \sigma_z \otimes \hat S'_1 \rangle &= -|\gamma| V \sin\left(\frac{\Delta\phi}{2}\right)\,\sin(\phi_\beta^{(1)} - \bar{\phi})&\nonumber \\
\langle \sigma_z \otimes \hat S'_3 \rangle &= |\gamma| V \sin\left(\frac{\Delta\phi}{2}\right)\cos(\phi_\beta^{(3)} - \bar{\phi})\nonumber \\
\langle \sigma_x \otimes \hat S'_1 \rangle &= |\gamma| V \cos\left(\phi + \frac{\Delta\phi}{2}\right) \cos(\phi_\beta^{(1)} - \bar{\phi})\nonumber \\
\langle \sigma_x \otimes \hat S'_2 \rangle &= |\gamma| V \sin\left(\phi + \frac{\Delta\phi}{2}\right) \sin\left(\frac{\Delta\phi}{2}\right)\nonumber \\
\langle \sigma_x \otimes \hat S'_3 \rangle &= |\gamma| V \cos\left(\phi+ \frac{\Delta\phi}{2}\right) \sin(\phi_\beta^{(3)} -\bar{\phi})\nonumber \\
\langle \sigma_y \otimes \hat S'_1 \rangle &= |\gamma| V \sin\left(\phi + \frac{\Delta\phi}{2}\right) \cos(\phi_\beta^{(1)} - \bar{\phi})\nonumber \\
\langle \sigma_y \otimes \hat S'_2 \rangle &= -|\gamma| V \sin\left(\frac{\Delta\phi}{2}\right) \cos\left(\phi + \frac{\Delta\phi}{2}\right)\nonumber \\
\langle \sigma_y \otimes \hat S'_3 \rangle &= |\gamma| V \sin\left(\phi + \frac{\Delta\phi}{2}\right) \sin(\phi_\beta^{(3)} - \bar{\phi}) 
\end{eqnarray}

As mentioned before, each correlation function is measured in a separate experimental run with a specific LO phase to target the orthogonal components of the macroscopic interference:
\begin{equation}
\phi_\beta^{(1)} = \bar{\phi}, \quad \phi_\beta^{(3)} = \bar{\phi} - \frac{\pi}{2} 
\end{equation}
\begin{itemize}
\item 

$S_1'$: in-phase interference component (real part of the coherence).
\item 

$S'_3$: orthogonal interference component (imaginary part of the coherence).
\end{itemize}
The correlation functions proportional to $\sin(\Delta\phi/2)$ are suppressed in the considered regime and do not contribute significantly. After optimizing the measurement via appropriate choice of the LO phase, for $V=1$, the surviving correlation functions are

\begin{eqnarray}
\label{calcorF}
\langle \sigma_x \otimes \hat S'_1 \rangle &=& |\gamma| \cos\phi\nonumber\\
\langle \sigma_x \otimes \hat S'_3 \rangle &=& -|\gamma| \cos\phi\nonumber\\
\langle \sigma_y \otimes \hat S'_1 \rangle &=& |\gamma| \sin\phi\nonumber\\
\langle \sigma_y \otimes \hat S'_3 \rangle &=& -|\gamma| \sin\phi
\end{eqnarray}
In this configuration, the photon-number difference measured at the output ports, $\hat{S}_1(\phi_\beta)$, serves as a phase-sensitive probe to reconstruct the correlations derived for the input modes. By toggling the LO phase between $\phi_\beta^{(1)}$ and $\phi_\beta^{(3)}$, the differential signal at the detectors captures the real and imaginary components of the macroscopic interference, respectively. This allows the measurement at the output ports to effectively map onto the mathematical input-mode operators $\hat{S}_1'$ and $\hat{S}_3'$, providing the data necessary to evaluate the bipartite witness.
\section{Derivation of the Entanglement Witness for a Separable State}
\label{appD}

{\it Statement}:
In the limit $|\gamma|=1$, the reconstructed entanglement witness $\langle \Pi^{T_M}_- \rangle_{\text{meas}}$ is non-negative for all separable states $\rho_{\text{sep}}$ generated within the considered mass-photon model.

{\it Proof}:
A general separable state in the bipartite Hilbert space is defined as a convex combination of product states:
\begin{equation}
    \rho_{\mathrm{sep}} = \sum_k p_k (\rho^k_m \otimes \rho^k_{\text{ph}}),
\end{equation}
where $\sum_k p_k = 1$ ($p_k \geq 0$), and $\rho^k_m$ and $\rho^k_{\text{ph}}$ denote the local density matrices for the matter and photon subsystems, respectively. For such states, joint expectation values of hybrid observables decompose into convex combinations of products of local expectations.

In this framework, the massive particle is treated as an effective qubit within the two-dimensional subspace spanned by the orthonormal spin-spatial states $\{|l, \uparrow\rangle, |r, \downarrow\rangle\}$~(see Eqs.\eqref{eq:deltaxlr}, and \eqref{eq:deltaxlr1}).~Within this reduced manifold, the matter observables—obtained by measuring the particle's spin—are mapped to Pauli operators, ensuring $|\langle \sigma_i \rangle_k| \leq 1$. Simultaneously, we define the normalized Stokes quantities operationally via the ratio of expectation values:
\begin{equation} \label{ratioDef}
    \langle \hat{S}'_j \rangle = \frac{\langle \hat{S}_j \rangle}{\langle \hat{S}_0 \rangle}.
\end{equation}
In the uncoupled limit ($\Delta\phi = 0$, $|\gamma|=1$, $\phi=0$)~(see Eqs.~\eqref{state}, \eqref{overlap}, \eqref{strength}, and \eqref{phase}), where $\alpha'=1$~\eqref{alphaprime}, the only correlation function that survives in the witness expression is $\langle \sigma_x \otimes \hat{S}'_1 \rangle$. Consequently, the measured witness~\eqref{Tmeas} reduces to:
\begin{equation}
    \langle \Pi^{T_M}_- \rangle_{\mathrm{meas}} = \frac{1}{4} \left( 1 - \frac{\sum_k p_k \langle \sigma_x \rangle_k \langle \hat{S}_1 \rangle_k}{\sum_k p_k \langle \hat{S}_0 \rangle_k} \right).
\end{equation}
For each product state $k$, the local expectations satisfy $|\langle \sigma_x \rangle_k| \leq 1$ and $|\langle \hat{S}_1 \rangle_k| \leq \langle \hat{S}_0 \rangle_k$. Therefore, we have the inequality:
\begin{equation}
    |\langle \sigma_x \rangle_k \langle \hat{S}_1 \rangle_k| \leq |\langle \hat{S}_1 \rangle_k| \leq \langle \hat{S}_0 \rangle_k,
\end{equation}
which implies:
\begin{equation}
    -\langle \hat{S}_0 \rangle_k \le \langle \sigma_x \rangle_k \langle \hat{S}_1 \rangle_k \le \langle \hat{S}_0 \rangle_k.
\end{equation}
Summing over all $k$ weighted by $p_k$, we obtain:
\begin{equation}
    \sum_k p_k \langle \sigma_x \rangle_k \langle \hat{S}_1 \rangle_k \leq \sum_k p_k \langle \hat{S}_0 \rangle_k.
\end{equation}
Since $\langle \hat{S}_0 \rangle_k \ge 0$, the denominator is non-negative (and strictly positive for nonzero optical intensity), ensuring:
\begin{equation}
    \frac{\sum_k p_k \langle \sigma_x \rangle_k \langle \hat{S}_1 \rangle_k}{\sum_k p_k \langle \hat{S}_0 \rangle_k} \leq 1.
\end{equation}
This guarantees that $\langle \Pi^{T_M}_- \rangle_{\mathrm{meas}} \geq 0$ for all separable states. Thus, any negative value of the witness certifies entanglement between the matter spin and the macroscopic optical field.

\section{Entanglement against isotropic noise}
\label{appE}
Adding isotropic noise gives
~~\(\rho_v = v~\rho + \frac{1-v}{4}~\openone_4~~\cite{Werner1989},\) where $v\in [0,1]$ is the noise parameter, \(\rho\) is the density matrix presented in Eq.~\eqref{matrix}, and \(\openone_4\) denotes the identity operator for the four-dimensional bipartite Hilbert space. After taking the partial transpose with respect to the massive particle subsystem, the eigenvalues of the resulting matrix $\rho_v^{T_M}$ are
\begin{align*}
\lambda_{1,2} &= \frac14\bigl(1+v \pm 2v|
\gamma|\bigr),~~\lambda_{3,4}= \frac14\bigl(1-v \pm 2v\sqrt{1-|
\gamma|^{2}}\bigr),
\end{align*}

\begin{widetext}
where
\[
\rho_v^{T_M}
=
\frac12
\begin{pmatrix}
v + \dfrac{1-v}{2} & 0 & v\,\gamma & 0 \\[6pt]
0 & \dfrac{1-v}{2} & v\,\sqrt{1-|\gamma|^{2}} & 0 \\[6pt]
v\,\gamma^{*} & v\,\sqrt{1-|\gamma|^{2}} &
v\,|\gamma|^{2} + \dfrac{1-v}{2} &
v\,\gamma\,\sqrt{1-|\gamma|^{2}} \\[6pt]
0 & 0 & v\,\gamma^{*}\sqrt{1-|\gamma|^{2}} &
v(1-|\gamma|^{2}) + \dfrac{1-v}{2}
\end{pmatrix}.
\]
\end{widetext}

The only eigenvalue that can become negative is
\(\lambda_{\mathrm{neg}} = \frac14\Bigl(1 - v\Bigr) - \frac{v}{2}\sqrt{1-|
\gamma|^{2}}\). For \(v>v_{\text{critical}},\) the state \(\rho_v\) is entangled. We obtain, 
\begin{equation}
\label{visibility}
   v_{\text{critical}}=\frac{1}{1+2\sqrt{1-|\gamma|^{2}}}.
\end{equation}

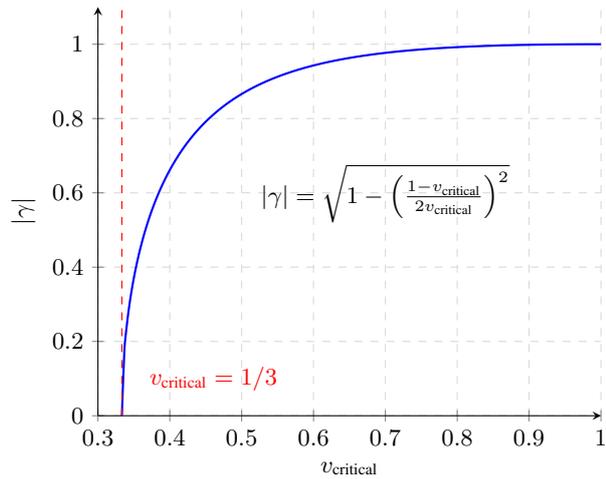
\begin{figure}[t]
    \centering
    \begin{tikzpicture}
        \begin{axis}[
            xlabel={$v_{\text{critical}}$},
            ylabel={$|\gamma|$},
            xmin=0.3, xmax=1.0,
            ymin=0, ymax=1.1,
            grid=major,
            grid style={dashed, gray!30},
            axis lines=left,
            samples=200,
            domain=0.333333:1, 
            width=8.2 cm, height=7cm]
            
            \addplot [
                blue, 
                thick,
                smooth
            ] {sqrt(max(0, 1 - ((1 - x)/(2*x))^2))};
            
            \node[anchor=north, black] at (axis cs:0.7, 0.7) 
            {$|\gamma| = \sqrt{1 - \left(\frac{1-v_{\text{critical}}}{2v_{\text{critical}}}\right)^2}$};
            
            \draw[red, dashed] (axis cs:0.333333, 0) -- (axis cs:0.333333, 1.1);
            \node[red, anchor=west] at (axis cs:0.36, 0.1) {$v_{\text{critical}} = 1/3$};
            
        \end{axis}
    \end{tikzpicture}
    \caption{Plot of $|\gamma|$ vs $v_{\text{critical}},$~(see Eq.~\eqref{visibility}) showing the dependence of the overlap parameter $|\gamma|$ on the noise parameter $v$. For any pure state of the mass-photon combined system $\eqref{matrix}$ (or any bipartite system) mixed with the isotropic noise, the minimum critical noise parameter $v_{\text{critical}}$ is $\frac{1}{3}.$} 
\end{figure}
Note that, the lowest value $v_{\text{critical}} = 1/3$ represents the absolute bound for the noise parameter when a bipartite pure state is mixed with isotropic noise. As we increase the overlap $|\gamma|$ (and thus the indistinguishability between the two states of the optical field), the required noise parameter $v$ to maintain entanglement increases significantly. In the limit of very high $|\gamma| \to 1$, the critical noise parameter $v_{\text{critical}}$ tends toward $1$, meaning even a trace amount of noise destroys the entanglement. Conversely, for $|\gamma|=0$, where the two states of the field are completely distinguishable, the bipartite pure state is most robust against isotropic noise. This behavior confirms that maximizing the distinguishability of the subsystems is the optimal strategy for protecting entanglement in a noisy environment.
\clearpage
\bibliography{reference.bib}

\begin{thebibliography}{58}%
\makeatletter
\providecommand \@ifxundefined [1]{%
 \@ifx{#1\undefined}
}%
\providecommand \@ifnum [1]{%
 \ifnum #1\expandafter \@firstoftwo
 \else \expandafter \@secondoftwo
 \fi
}%
\providecommand \@ifx [1]{%
 \ifx #1\expandafter \@firstoftwo
 \else \expandafter \@secondoftwo
 \fi
}%
\providecommand \natexlab [1]{#1}%
\providecommand \enquote  [1]{``#1''}%
\providecommand \bibnamefont  [1]{#1}%
\providecommand \bibfnamefont [1]{#1}%
\providecommand \citenamefont [1]{#1}%
\providecommand \href@noop [0]{\@secondoftwo}%
\providecommand \href [0]{\begingroup \@sanitize@url \@href}%
\providecommand \@href[1]{\@@startlink{#1}\@@href}%
\providecommand \@@href[1]{\endgroup#1\@@endlink}%
\providecommand \@sanitize@url [0]{\catcode `\\12\catcode `\$12\catcode `\&12\catcode `\#12\catcode `\^12\catcode `\_12\catcode `\%12\relax}%
\providecommand \@@startlink[1]{}%
\providecommand \@@endlink[0]{}%
\providecommand \url  [0]{\begingroup\@sanitize@url \@url }%
\providecommand \@url [1]{\endgroup\@href {#1}{\urlprefix }}%
\providecommand \urlprefix  [0]{URL }%
\providecommand \Eprint [0]{\href }%
\providecommand \doibase [0]{https://doi.org/}%
\providecommand \selectlanguage [0]{\@gobble}%
\providecommand \bibinfo  [0]{\@secondoftwo}%
\providecommand \bibfield  [0]{\@secondoftwo}%
\providecommand \translation [1]{[#1]}%
\providecommand \BibitemOpen [0]{}%
\providecommand \bibitemStop [0]{}%
\providecommand \bibitemNoStop [0]{.\EOS\space}%
\providecommand \EOS [0]{\spacefactor3000\relax}%
\providecommand \BibitemShut  [1]{\csname bibitem#1\endcsname}%
\let\auto@bib@innerbib\@empty
\bibitem [{\citenamefont {Bose}\ \emph {et~al.}(2025)\citenamefont {Bose} \emph {et~al.}}]{Bose:2025qns}%
  \BibitemOpen
  \bibfield  {author} {\bibinfo {author} {\bibfnamefont {S.}~\bibnamefont {Bose}} \emph {et~al.},\ }\bibfield  {title} {\bibinfo {title} {{A Spin-Based Pathway to Testing the Quantum Nature of Gravity}}\ }(\bibinfo {year} {2025})\ \Eprint {https://arxiv.org/abs/2509.01586} {arXiv:2509.01586 [quant-ph]} \BibitemShut {NoStop}%
\bibitem [{\citenamefont {Bose}\ \emph {et~al.}(2017)\citenamefont {Bose}, \citenamefont {Mazumdar}, \citenamefont {Morley}, \citenamefont {Ulbricht}, \citenamefont {Toro{\v{s}}}, \citenamefont {Paternostro}, \citenamefont {Geraci}, \citenamefont {Barker}, \citenamefont {Kim},\ and\ \citenamefont {Milburn}}]{Bose:2017nin}%
  \BibitemOpen
  \bibfield  {author} {\bibinfo {author} {\bibfnamefont {S.}~\bibnamefont {Bose}}, \bibinfo {author} {\bibfnamefont {A.}~\bibnamefont {Mazumdar}}, \bibinfo {author} {\bibfnamefont {G.~W.}\ \bibnamefont {Morley}}, \bibinfo {author} {\bibfnamefont {H.}~\bibnamefont {Ulbricht}}, \bibinfo {author} {\bibfnamefont {M.}~\bibnamefont {Toro{\v{s}}}}, \bibinfo {author} {\bibfnamefont {M.}~\bibnamefont {Paternostro}}, \bibinfo {author} {\bibfnamefont {A.}~\bibnamefont {Geraci}}, \bibinfo {author} {\bibfnamefont {P.}~\bibnamefont {Barker}}, \bibinfo {author} {\bibfnamefont {M.~S.}\ \bibnamefont {Kim}},\ and\ \bibinfo {author} {\bibfnamefont {G.}~\bibnamefont {Milburn}},\ }\bibfield  {title} {\bibinfo {title} {{Spin Entanglement Witness for Quantum Gravity}},\ }\href {https://doi.org/10.1103/PhysRevLett.119.240401} {\bibfield  {journal} {\bibinfo  {journal} {Phys. Rev. Lett.}\ }\textbf {\bibinfo {volume} {119}},\ \bibinfo {pages} {240401} (\bibinfo {year} {2017})},\ \Eprint {https://arxiv.org/abs/1707.06050}
  {arXiv:1707.06050 [quant-ph]} \BibitemShut {NoStop}%
\bibitem [{ICT(2016)}]{ICTS}%
  \BibitemOpen
  \href@noop {} {}\bibinfo {howpublished} {\url{https://www.youtube.com/watch?v=0Fv-0k13s_k}} (\bibinfo {year} {2016}),\ \bibinfo {note} {accessed 1/11/22}\BibitemShut {NoStop}%
\bibitem [{\citenamefont {Elahi}\ and\ \citenamefont {Mazumdar}(2023)}]{Elahi:2023ozf}%
  \BibitemOpen
  \bibfield  {author} {\bibinfo {author} {\bibfnamefont {S.~G.}\ \bibnamefont {Elahi}}\ and\ \bibinfo {author} {\bibfnamefont {A.}~\bibnamefont {Mazumdar}},\ }\bibfield  {title} {\bibinfo {title} {{Probing massless and massive gravitons via entanglement in a warped extra dimension}},\ }\href {https://doi.org/10.1103/PhysRevD.108.035018} {\bibfield  {journal} {\bibinfo  {journal} {Phys. Rev. D}\ }\textbf {\bibinfo {volume} {108}},\ \bibinfo {pages} {035018} (\bibinfo {year} {2023})},\ \Eprint {https://arxiv.org/abs/2303.07371} {arXiv:2303.07371 [gr-qc]} \BibitemShut {NoStop}%
\bibitem [{\citenamefont {Marletto}\ and\ \citenamefont {Vedral}(2017)}]{Marletto:2017kzi}%
  \BibitemOpen
  \bibfield  {author} {\bibinfo {author} {\bibfnamefont {C.}~\bibnamefont {Marletto}}\ and\ \bibinfo {author} {\bibfnamefont {V.}~\bibnamefont {Vedral}},\ }\bibfield  {title} {\bibinfo {title} {{Gravitationally-induced entanglement between two massive particles is sufficient evidence of quantum effects in gravity}},\ }\href {https://doi.org/10.1103/PhysRevLett.119.240402} {\bibfield  {journal} {\bibinfo  {journal} {Phys. Rev. Lett.}\ }\textbf {\bibinfo {volume} {119}},\ \bibinfo {pages} {240402} (\bibinfo {year} {2017})},\ \Eprint {https://arxiv.org/abs/1707.06036} {arXiv:1707.06036 [quant-ph]} \BibitemShut {NoStop}%
\bibitem [{\citenamefont {Marshman}\ \emph {et~al.}(2020{\natexlab{a}})\citenamefont {Marshman}, \citenamefont {Mazumdar},\ and\ \citenamefont {Bose}}]{Marshman:2019sne}%
  \BibitemOpen
  \bibfield  {author} {\bibinfo {author} {\bibfnamefont {R.~J.}\ \bibnamefont {Marshman}}, \bibinfo {author} {\bibfnamefont {A.}~\bibnamefont {Mazumdar}},\ and\ \bibinfo {author} {\bibfnamefont {S.}~\bibnamefont {Bose}},\ }\bibfield  {title} {\bibinfo {title} {{Locality and entanglement in table-top testing of the quantum nature of linearized gravity}},\ }\href {https://doi.org/10.1103/PhysRevA.101.052110} {\bibfield  {journal} {\bibinfo  {journal} {Phys. Rev. A}\ }\textbf {\bibinfo {volume} {101}},\ \bibinfo {pages} {052110} (\bibinfo {year} {2020}{\natexlab{a}})},\ \Eprint {https://arxiv.org/abs/1907.01568} {arXiv:1907.01568 [quant-ph]} \BibitemShut {NoStop}%
\bibitem [{\citenamefont {Bose}\ \emph {et~al.}(2022)\citenamefont {Bose}, \citenamefont {Mazumdar}, \citenamefont {Schut},\ and\ \citenamefont {Toro\v{s}}}]{Bose:2022uxe}%
  \BibitemOpen
  \bibfield  {author} {\bibinfo {author} {\bibfnamefont {S.}~\bibnamefont {Bose}}, \bibinfo {author} {\bibfnamefont {A.}~\bibnamefont {Mazumdar}}, \bibinfo {author} {\bibfnamefont {M.}~\bibnamefont {Schut}},\ and\ \bibinfo {author} {\bibfnamefont {M.}~\bibnamefont {Toro\v{s}}},\ }\bibfield  {title} {\bibinfo {title} {{Mechanism for the quantum natured gravitons to entangle masses}},\ }\href {https://doi.org/10.1103/PhysRevD.105.106028} {\bibfield  {journal} {\bibinfo  {journal} {Phys. Rev. D}\ }\textbf {\bibinfo {volume} {105}},\ \bibinfo {pages} {106028} (\bibinfo {year} {2022})},\ \Eprint {https://arxiv.org/abs/2201.03583} {arXiv:2201.03583 [gr-qc]} \BibitemShut {NoStop}%
\bibitem [{\citenamefont {Beckering~Vinckers}\ \emph {et~al.}(2023)\citenamefont {Beckering~Vinckers}, \citenamefont {De~La Cruz-Dombriz},\ and\ \citenamefont {Mazumdar}}]{Vinckers:2023grv}%
  \BibitemOpen
  \bibfield  {author} {\bibinfo {author} {\bibfnamefont {U.~K.}\ \bibnamefont {Beckering~Vinckers}}, \bibinfo {author} {\bibfnamefont {{\'A}.}~\bibnamefont {De~La Cruz-Dombriz}},\ and\ \bibinfo {author} {\bibfnamefont {A.}~\bibnamefont {Mazumdar}},\ }\bibfield  {title} {\bibinfo {title} {Quantum entanglement of masses with nonlocal gravitational interaction},\ }\href {https://doi.org/10.1103/PhysRevD.107.124036} {\bibfield  {journal} {\bibinfo  {journal} {Physical Review D}\ }\textbf {\bibinfo {volume} {107}},\ \bibinfo {pages} {124036} (\bibinfo {year} {2023})}\BibitemShut {NoStop}%
\bibitem [{\citenamefont {Dyson}\ \emph {et~al.}(1920)\citenamefont {Dyson}, \citenamefont {Eddington},\ and\ \citenamefont {Davidson}}]{Eddington}%
  \BibitemOpen
  \bibfield  {author} {\bibinfo {author} {\bibfnamefont {F.~W.}\ \bibnamefont {Dyson}}, \bibinfo {author} {\bibfnamefont {A.~S.}\ \bibnamefont {Eddington}},\ and\ \bibinfo {author} {\bibfnamefont {C.}~\bibnamefont {Davidson}},\ }\bibfield  {title} {\bibinfo {title} {A determination of the deflection of light by the sun's gravitational field, from observations made at the total eclipse of may 29, 1919},\ }\href {https://doi.org/doi.org/10.1098/rsta.1920.0009} {\bibfield  {journal} {\bibinfo  {journal} {Philos Trans A Math Phys Eng Sci 1}\ }\textbf {\bibinfo {volume} {220}},\ \bibinfo {pages} {291} (\bibinfo {year} {1920})}\BibitemShut {NoStop}%
\bibitem [{\citenamefont {Nordstr{\"o}m}(1913)}]{Nordstrom:1913dga}%
  \BibitemOpen
  \bibfield  {author} {\bibinfo {author} {\bibfnamefont {G.}~\bibnamefont {Nordstr{\"o}m}},\ }\bibfield  {title} {\bibinfo {title} {{Zur Theorie der Gravitation vom Standpunkt des Relativit{\"a}tsprinzips}},\ }\href {https://doi.org/10.1002/andp.19133471303} {\bibfield  {journal} {\bibinfo  {journal} {Annalen Phys.}\ }\textbf {\bibinfo {volume} {347}},\ \bibinfo {pages} {533} (\bibinfo {year} {1913})}\BibitemShut {NoStop}%
\bibitem [{\citenamefont {Biswas}\ \emph {et~al.}(2012)\citenamefont {Biswas}, \citenamefont {Gerwick}, \citenamefont {Koivisto},\ and\ \citenamefont {Mazumdar}}]{Biswas:2011ar}%
  \BibitemOpen
  \bibfield  {author} {\bibinfo {author} {\bibfnamefont {T.}~\bibnamefont {Biswas}}, \bibinfo {author} {\bibfnamefont {E.}~\bibnamefont {Gerwick}}, \bibinfo {author} {\bibfnamefont {T.}~\bibnamefont {Koivisto}},\ and\ \bibinfo {author} {\bibfnamefont {A.}~\bibnamefont {Mazumdar}},\ }\bibfield  {title} {\bibinfo {title} {{Towards singularity and ghost free theories of gravity}},\ }\href {https://doi.org/10.1103/PhysRevLett.108.031101} {\bibfield  {journal} {\bibinfo  {journal} {Phys. Rev. Lett.}\ }\textbf {\bibinfo {volume} {108}},\ \bibinfo {pages} {031101} (\bibinfo {year} {2012})},\ \Eprint {https://arxiv.org/abs/1110.5249} {arXiv:1110.5249 [gr-qc]} \BibitemShut {NoStop}%
\bibitem [{\citenamefont {Biswas}\ \emph {et~al.}(2013)\citenamefont {Biswas}, \citenamefont {Koivisto},\ and\ \citenamefont {Mazumdar}}]{Biswas:2013kla}%
  \BibitemOpen
  \bibfield  {author} {\bibinfo {author} {\bibfnamefont {T.}~\bibnamefont {Biswas}}, \bibinfo {author} {\bibfnamefont {T.}~\bibnamefont {Koivisto}},\ and\ \bibinfo {author} {\bibfnamefont {A.}~\bibnamefont {Mazumdar}},\ }\bibfield  {title} {\bibinfo {title} {{Nonlocal theories of gravity: the flat space propagator}},\ }in\ \href@noop {} {\emph {\bibinfo {booktitle} {{Barcelona Postgrad Encounters on Fundamental Physics}}}}\ (\bibinfo {year} {2013})\ pp.\ \bibinfo {pages} {13--24},\ \Eprint {https://arxiv.org/abs/1302.0532} {arXiv:1302.0532 [gr-qc]} \BibitemShut {NoStop}%
\bibitem [{\citenamefont {Rufo}\ \emph {et~al.}(2025)\citenamefont {Rufo}, \citenamefont {Mazumdar},\ and\ \citenamefont {Sab{\'\i}n}}]{Rufo:2024ulr}%
  \BibitemOpen
  \bibfield  {author} {\bibinfo {author} {\bibfnamefont {P.~G.~C.}\ \bibnamefont {Rufo}}, \bibinfo {author} {\bibfnamefont {A.}~\bibnamefont {Mazumdar}},\ and\ \bibinfo {author} {\bibfnamefont {C.}~\bibnamefont {Sab{\'\i}n}},\ }\bibfield  {title} {\bibinfo {title} {{Genuine tripartite entanglement in graviton-matter interactions}},\ }\href {https://doi.org/10.1103/PhysRevA.111.022444} {\bibfield  {journal} {\bibinfo  {journal} {Phys. Rev. A}\ }\textbf {\bibinfo {volume} {111}},\ \bibinfo {pages} {022444} (\bibinfo {year} {2025})},\ \Eprint {https://arxiv.org/abs/2411.03293} {arXiv:2411.03293 [quant-ph]} \BibitemShut {NoStop}%
\bibitem [{\citenamefont {Biswas}\ \emph {et~al.}(2023)\citenamefont {Biswas}, \citenamefont {Bose}, \citenamefont {Mazumdar},\ and\ \citenamefont {Toro{\v{s}}}}]{Biswas:2022qto}%
  \BibitemOpen
  \bibfield  {author} {\bibinfo {author} {\bibfnamefont {D.}~\bibnamefont {Biswas}}, \bibinfo {author} {\bibfnamefont {S.}~\bibnamefont {Bose}}, \bibinfo {author} {\bibfnamefont {A.}~\bibnamefont {Mazumdar}},\ and\ \bibinfo {author} {\bibfnamefont {M.}~\bibnamefont {Toro{\v{s}}}},\ }\bibfield  {title} {\bibinfo {title} {{Gravitational optomechanics: Photon-matter entanglement via graviton exchange}},\ }\href {https://doi.org/10.1103/PhysRevD.108.064023} {\bibfield  {journal} {\bibinfo  {journal} {Phys. Rev. D}\ }\textbf {\bibinfo {volume} {108}},\ \bibinfo {pages} {064023} (\bibinfo {year} {2023})},\ \Eprint {https://arxiv.org/abs/2209.09273} {arXiv:2209.09273 [gr-qc]} \BibitemShut {NoStop}%
\bibitem [{\citenamefont {Carney}(2022)}]{Carney:2021vvt}%
  \BibitemOpen
  \bibfield  {author} {\bibinfo {author} {\bibfnamefont {D.}~\bibnamefont {Carney}},\ }\bibfield  {title} {\bibinfo {title} {{Newton, entanglement, and the graviton}},\ }\href {https://doi.org/10.1103/PhysRevD.105.024029} {\bibfield  {journal} {\bibinfo  {journal} {Phys. Rev. D}\ }\textbf {\bibinfo {volume} {105}},\ \bibinfo {pages} {024029} (\bibinfo {year} {2022})},\ \Eprint {https://arxiv.org/abs/2108.06320} {arXiv:2108.06320 [quant-ph]} \BibitemShut {NoStop}%
\bibitem [{\citenamefont {Scadron}(2007)}]{Scadron:2007qd}%
  \BibitemOpen
  \bibfield  {author} {\bibinfo {author} {\bibfnamefont {M.~D.}\ \bibnamefont {Scadron}},\ }\href@noop {} {\emph {\bibinfo {title} {{Advanced quantum theory}}}}\ (\bibinfo {year} {2007})\BibitemShut {NoStop}%
\bibitem [{\citenamefont {Schnabel}\ \emph {et~al.}(2003)\citenamefont {Schnabel}, \citenamefont {Bowen}, \citenamefont {Treps}, \citenamefont {Ralph}, \citenamefont {Bachor},\ and\ \citenamefont {Lam}}]{schnabel2003stokes}%
  \BibitemOpen
  \bibfield  {author} {\bibinfo {author} {\bibfnamefont {R.}~\bibnamefont {Schnabel}}, \bibinfo {author} {\bibfnamefont {W.~P.}\ \bibnamefont {Bowen}}, \bibinfo {author} {\bibfnamefont {N.}~\bibnamefont {Treps}}, \bibinfo {author} {\bibfnamefont {T.~C.}\ \bibnamefont {Ralph}}, \bibinfo {author} {\bibfnamefont {H.-A.}\ \bibnamefont {Bachor}},\ and\ \bibinfo {author} {\bibfnamefont {P.~K.}\ \bibnamefont {Lam}},\ }\bibfield  {title} {\bibinfo {title} {Stokes-operator-squeezed continuous-variable polarization states},\ }\href {https://doi.org/10.1103/PhysRevA.67.012316} {\bibfield  {journal} {\bibinfo  {journal} {Physical Review A}\ }\textbf {\bibinfo {volume} {67}},\ \bibinfo {pages} {012316} (\bibinfo {year} {2003})}\BibitemShut {NoStop}%
\bibitem [{\citenamefont {Agarwal}\ and\ \citenamefont {Chaturvedi}(2003)}]{Agarwal2003Scheme}%
  \BibitemOpen
  \bibfield  {author} {\bibinfo {author} {\bibfnamefont {G.~S.}\ \bibnamefont {Agarwal}}\ and\ \bibinfo {author} {\bibfnamefont {S.}~\bibnamefont {Chaturvedi}},\ }\bibfield  {title} {\bibinfo {title} {Scheme to measure quantum stokes parameters and their fluctuations and correlations},\ }\href {https://doi.org/10.1080/09500340308235179} {\bibfield  {journal} {\bibinfo  {journal} {Journal of Modern Optics}\ }\textbf {\bibinfo {volume} {50}},\ \bibinfo {pages} {711} (\bibinfo {year} {2003})}\BibitemShut {NoStop}%
\bibitem [{\citenamefont {Stokes}(1851)}]{stokes1851composition}%
  \BibitemOpen
  \bibfield  {author} {\bibinfo {author} {\bibfnamefont {G.~G.}\ \bibnamefont {Stokes}},\ }\bibfield  {title} {\bibinfo {title} {On the composition and resolution of streams of polarized light from different sources},\ }\href@noop {} {\bibfield  {journal} {\bibinfo  {journal} {Transactions of the Cambridge Philosophical Society}\ }\textbf {\bibinfo {volume} {9}},\ \bibinfo {pages} {399} (\bibinfo {year} {1851})}\BibitemShut {NoStop}%
\bibitem [{\citenamefont {Peres}(1996)}]{Peres:1996dw}%
  \BibitemOpen
  \bibfield  {author} {\bibinfo {author} {\bibfnamefont {A.}~\bibnamefont {Peres}},\ }\bibfield  {title} {\bibinfo {title} {{Separability criterion for density matrices}},\ }\href {https://doi.org/10.1103/PhysRevLett.77.1413} {\bibfield  {journal} {\bibinfo  {journal} {Phys. Rev. Lett.}\ }\textbf {\bibinfo {volume} {77}},\ \bibinfo {pages} {1413} (\bibinfo {year} {1996})},\ \Eprint {https://arxiv.org/abs/quant-ph/9604005} {arXiv:quant-ph/9604005} \BibitemShut {NoStop}%
\bibitem [{\citenamefont {Horodecki}\ \emph {et~al.}(1996)\citenamefont {Horodecki}, \citenamefont {Horodecki},\ and\ \citenamefont {Horodecki}}]{horodecki1996separability}%
  \BibitemOpen
  \bibfield  {author} {\bibinfo {author} {\bibfnamefont {M.}~\bibnamefont {Horodecki}}, \bibinfo {author} {\bibfnamefont {P.}~\bibnamefont {Horodecki}},\ and\ \bibinfo {author} {\bibfnamefont {R.}~\bibnamefont {Horodecki}},\ }\bibfield  {title} {\bibinfo {title} {Separability of mixed states: necessary and sufficient conditions},\ }\href {https://doi.org/10.1016/S0375-9601(96)00706-2} {\bibfield  {journal} {\bibinfo  {journal} {Physics Letters A}\ }\textbf {\bibinfo {volume} {223}},\ \bibinfo {pages} {1} (\bibinfo {year} {1996})}\BibitemShut {NoStop}%
\bibitem [{\citenamefont {Doherty}\ \emph {et~al.}(2013)\citenamefont {Doherty}, \citenamefont {Manson}, \citenamefont {Delaney}, \citenamefont {Jelezko}, \citenamefont {Wrachtrup},\ and\ \citenamefont {Hollenberg}}]{Doherty_2013}%
  \BibitemOpen
  \bibfield  {author} {\bibinfo {author} {\bibfnamefont {M.~W.}\ \bibnamefont {Doherty}}, \bibinfo {author} {\bibfnamefont {N.~B.}\ \bibnamefont {Manson}}, \bibinfo {author} {\bibfnamefont {P.}~\bibnamefont {Delaney}}, \bibinfo {author} {\bibfnamefont {F.}~\bibnamefont {Jelezko}}, \bibinfo {author} {\bibfnamefont {J.}~\bibnamefont {Wrachtrup}},\ and\ \bibinfo {author} {\bibfnamefont {L.~C.}\ \bibnamefont {Hollenberg}},\ }\bibfield  {title} {\bibinfo {title} {The nitrogen-vacancy colour centre in diamond},\ }\href {https://doi.org/10.1016/j.physrep.2013.02.001} {\bibfield  {journal} {\bibinfo  {journal} {Physics Reports}\ }\textbf {\bibinfo {volume} {528}},\ \bibinfo {pages} {1–45} (\bibinfo {year} {2013})}\BibitemShut {NoStop}%
\bibitem [{\citenamefont {Korolkova}\ and\ \citenamefont {Chirkin}(1996)}]{Korolkova1996}%
  \BibitemOpen
  \bibfield  {author} {\bibinfo {author} {\bibfnamefont {N.~V.}\ \bibnamefont {Korolkova}}\ and\ \bibinfo {author} {\bibfnamefont {A.~S.}\ \bibnamefont {Chirkin}},\ }\bibfield  {title} {\bibinfo {title} {Polarization squeezing and photon-number correlations in a periodically nonlinear medium},\ }\href@noop {} {\bibfield  {journal} {\bibinfo  {journal} {Journal of Modern Optics}\ }\textbf {\bibinfo {volume} {43}},\ \bibinfo {pages} {869} (\bibinfo {year} {1996})}\BibitemShut {NoStop}%
\bibitem [{\citenamefont {Korolkova}\ \emph {et~al.}(2002)\citenamefont {Korolkova}, \citenamefont {Leuchs}, \citenamefont {Loudon}, \citenamefont {Ralph},\ and\ \citenamefont {Silberhorn}}]{korolkova2002polarization}%
  \BibitemOpen
  \bibfield  {author} {\bibinfo {author} {\bibfnamefont {N.}~\bibnamefont {Korolkova}}, \bibinfo {author} {\bibfnamefont {G.}~\bibnamefont {Leuchs}}, \bibinfo {author} {\bibfnamefont {R.}~\bibnamefont {Loudon}}, \bibinfo {author} {\bibfnamefont {T.~C.}\ \bibnamefont {Ralph}},\ and\ \bibinfo {author} {\bibfnamefont {C.}~\bibnamefont {Silberhorn}},\ }\bibfield  {title} {\bibinfo {title} {Polarization squeezing and continuous-variable polarization entanglement},\ }\href {https://doi.org/10.1103/PhysRevA.65.052306} {\bibfield  {journal} {\bibinfo  {journal} {Physical Review A}\ }\textbf {\bibinfo {volume} {65}},\ \bibinfo {pages} {052306} (\bibinfo {year} {2002})}\BibitemShut {NoStop}%
\bibitem [{\citenamefont {\ifmmode~\dot{Z}\else \.{Z}\fi{}ukowski}\ \emph {et~al.}(2017)\citenamefont {\ifmmode~\dot{Z}\else \.{Z}\fi{}ukowski}, \citenamefont {Laskowski},\ and\ \citenamefont {Wie\ifmmode~\acute{s}\else \'{s}\fi{}niak}}]{zukowski}%
  \BibitemOpen
  \bibfield  {author} {\bibinfo {author} {\bibfnamefont {M.}~\bibnamefont {\ifmmode~\dot{Z}\else \.{Z}\fi{}ukowski}}, \bibinfo {author} {\bibfnamefont {W.}~\bibnamefont {Laskowski}},\ and\ \bibinfo {author} {\bibfnamefont {M.}~\bibnamefont {Wie\ifmmode~\acute{s}\else \'{s}\fi{}niak}},\ }\bibfield  {title} {\bibinfo {title} {Normalized stokes operators for polarization correlations of entangled optical fields},\ }\href {https://doi.org/10.1103/PhysRevA.95.042113} {\bibfield  {journal} {\bibinfo  {journal} {Phys. Rev. A}\ }\textbf {\bibinfo {volume} {95}},\ \bibinfo {pages} {042113} (\bibinfo {year} {2017})}\BibitemShut {NoStop}%
\bibitem [{\citenamefont {Zhou}\ \emph {et~al.}(2023)\citenamefont {Zhou}, \citenamefont {Marshman}, \citenamefont {Bose},\ and\ \citenamefont {Mazumdar}}]{Zhou:2022jug}%
  \BibitemOpen
  \bibfield  {author} {\bibinfo {author} {\bibfnamefont {R.}~\bibnamefont {Zhou}}, \bibinfo {author} {\bibfnamefont {R.~J.}\ \bibnamefont {Marshman}}, \bibinfo {author} {\bibfnamefont {S.}~\bibnamefont {Bose}},\ and\ \bibinfo {author} {\bibfnamefont {A.}~\bibnamefont {Mazumdar}},\ }\bibfield  {title} {\bibinfo {title} {{Mass-independent scheme for enhancing spatial quantum superpositions}},\ }\href {https://doi.org/10.1103/PhysRevA.107.032212} {\bibfield  {journal} {\bibinfo  {journal} {Phys. Rev. A}\ }\textbf {\bibinfo {volume} {107}},\ \bibinfo {pages} {032212} (\bibinfo {year} {2023})},\ \Eprint {https://arxiv.org/abs/2210.05689} {arXiv:2210.05689 [quant-ph]} \BibitemShut {NoStop}%
\bibitem [{\citenamefont {Zhou}\ \emph {et~al.}(2022)\citenamefont {Zhou}, \citenamefont {Marshman}, \citenamefont {Bose},\ and\ \citenamefont {Mazumdar}}]{Zhou:2022frl}%
  \BibitemOpen
  \bibfield  {author} {\bibinfo {author} {\bibfnamefont {R.}~\bibnamefont {Zhou}}, \bibinfo {author} {\bibfnamefont {R.~J.}\ \bibnamefont {Marshman}}, \bibinfo {author} {\bibfnamefont {S.}~\bibnamefont {Bose}},\ and\ \bibinfo {author} {\bibfnamefont {A.}~\bibnamefont {Mazumdar}},\ }\bibfield  {title} {\bibinfo {title} {{Catapulting towards massive and large spatial quantum superposition}},\ }\href {https://doi.org/10.1103/PhysRevResearch.4.043157} {\bibfield  {journal} {\bibinfo  {journal} {Phys. Rev. Res.}\ }\textbf {\bibinfo {volume} {4}},\ \bibinfo {pages} {043157} (\bibinfo {year} {2022})},\ \Eprint {https://arxiv.org/abs/2206.04088} {arXiv:2206.04088 [quant-ph]} \BibitemShut {NoStop}%
\bibitem [{\citenamefont {Zhou}\ \emph {et~al.}(2024)\citenamefont {Zhou}, \citenamefont {Marshman},\ and\ \citenamefont {Bose}}]{Zhou:2022epb}%
  \BibitemOpen
  \bibfield  {author} {\bibinfo {author} {\bibfnamefont {R.}~\bibnamefont {Zhou}}, \bibinfo {author} {\bibfnamefont {R.~J.}\ \bibnamefont {Marshman}},\ and\ \bibinfo {author} {\bibfnamefont {S.}~\bibnamefont {Bose}},\ }\bibfield  {title} {\bibinfo {title} {Gravito-diamagnetic forces for mass independent large spatial superpositions},\ }\href {https://doi.org/10.1088/1402-4896/ad37df} {\bibfield  {journal} {\bibinfo  {journal} {Physica Scripta}\ }\textbf {\bibinfo {volume} {99}},\ \bibinfo {pages} {055114} (\bibinfo {year} {2024})},\ \bibinfo {note} {open Access}\BibitemShut {NoStop}%
\bibitem [{\citenamefont {Zhou}\ \emph {et~al.}(2025)\citenamefont {Zhou}, \citenamefont {Xiang},\ and\ \citenamefont {Mazumdar}}]{Zhou:2024voj}%
  \BibitemOpen
  \bibfield  {author} {\bibinfo {author} {\bibfnamefont {R.}~\bibnamefont {Zhou}}, \bibinfo {author} {\bibfnamefont {Q.}~\bibnamefont {Xiang}},\ and\ \bibinfo {author} {\bibfnamefont {A.}~\bibnamefont {Mazumdar}},\ }\bibfield  {title} {\bibinfo {title} {{Spin-dependent force and inverted harmonic potential for rapid creation of macroscopic quantum superpositions}},\ }\href {https://doi.org/10.1103/PhysRevA.111.052207} {\bibfield  {journal} {\bibinfo  {journal} {Phys. Rev. A}\ }\textbf {\bibinfo {volume} {111}},\ \bibinfo {pages} {052207} (\bibinfo {year} {2025})},\ \Eprint {https://arxiv.org/abs/2408.11909} {arXiv:2408.11909 [quant-ph]} \BibitemShut {NoStop}%
\bibitem [{\citenamefont {Marshman}\ \emph {et~al.}(2020{\natexlab{b}})\citenamefont {Marshman}, \citenamefont {Mazumdar}, \citenamefont {Morley}, \citenamefont {Barker}, \citenamefont {Hoekstra},\ and\ \citenamefont {Bose}}]{Marshman:2018upe}%
  \BibitemOpen
  \bibfield  {author} {\bibinfo {author} {\bibfnamefont {R.~J.}\ \bibnamefont {Marshman}}, \bibinfo {author} {\bibfnamefont {A.}~\bibnamefont {Mazumdar}}, \bibinfo {author} {\bibfnamefont {G.~W.}\ \bibnamefont {Morley}}, \bibinfo {author} {\bibfnamefont {P.~F.}\ \bibnamefont {Barker}}, \bibinfo {author} {\bibfnamefont {S.}~\bibnamefont {Hoekstra}},\ and\ \bibinfo {author} {\bibfnamefont {S.}~\bibnamefont {Bose}},\ }\bibfield  {title} {\bibinfo {title} {{Mesoscopic Interference for Metric and Curvature (MIMAC) $\&$ Gravitational Wave Detection}},\ }\href {https://doi.org/10.1088/1367-2630/ab9f6c} {\bibfield  {journal} {\bibinfo  {journal} {New J. Phys.}\ }\textbf {\bibinfo {volume} {22}},\ \bibinfo {pages} {083012} (\bibinfo {year} {2020}{\natexlab{b}})},\ \Eprint {https://arxiv.org/abs/1807.10830} {arXiv:1807.10830 [gr-qc]} \BibitemShut {NoStop}%
\bibitem [{\citenamefont {Hornberger}(2009)}]{Hornberger:2008xkz}%
  \BibitemOpen
  \bibfield  {author} {\bibinfo {author} {\bibfnamefont {K.}~\bibnamefont {Hornberger}},\ }\bibfield  {title} {\bibinfo {title} {{Introduction to decoherence theory}},\ }\href {https://doi.org/10.1007/978-3-540-88169-8_5} {\bibfield  {journal} {\bibinfo  {journal} {Lect. Notes Phys.}\ }\textbf {\bibinfo {volume} {768}},\ \bibinfo {pages} {221} (\bibinfo {year} {2009})},\ \Eprint {https://arxiv.org/abs/quant-ph/0612118} {arXiv:quant-ph/0612118} \BibitemShut {NoStop}%
\bibitem [{\citenamefont {Arndt}\ and\ \citenamefont {Hornberger}(2014)}]{Arndt:2014blv}%
  \BibitemOpen
  \bibfield  {author} {\bibinfo {author} {\bibfnamefont {M.}~\bibnamefont {Arndt}}\ and\ \bibinfo {author} {\bibfnamefont {K.}~\bibnamefont {Hornberger}},\ }\bibfield  {title} {\bibinfo {title} {{Testing the limits of quantum mechanical superpositions}},\ }\href {https://doi.org/10.1038/nphys2863} {\bibfield  {journal} {\bibinfo  {journal} {Nature Phys.}\ }\textbf {\bibinfo {volume} {10}},\ \bibinfo {pages} {271} (\bibinfo {year} {2014})},\ \Eprint {https://arxiv.org/abs/1410.0270} {arXiv:1410.0270 [quant-ph]} \BibitemShut {NoStop}%
\bibitem [{\citenamefont {Romero-Isart}(2011)}]{Romero-Isart:2011yun}%
  \BibitemOpen
  \bibfield  {author} {\bibinfo {author} {\bibfnamefont {O.}~\bibnamefont {Romero-Isart}},\ }\bibfield  {title} {\bibinfo {title} {{Quantum superposition of massive objects and collapse models}},\ }\href {https://doi.org/10.1103/PhysRevA.84.052121} {\bibfield  {journal} {\bibinfo  {journal} {Phys. Rev. A}\ }\textbf {\bibinfo {volume} {84}},\ \bibinfo {pages} {052121} (\bibinfo {year} {2011})},\ \Eprint {https://arxiv.org/abs/1110.4495} {arXiv:1110.4495 [quant-ph]} \BibitemShut {NoStop}%
\bibitem [{\citenamefont {Romero-Isart}\ \emph {et~al.}(2011)\citenamefont {Romero-Isart}, \citenamefont {Pflanzer}, \citenamefont {Blaser}, \citenamefont {Kaltenbaek}, \citenamefont {Kiesel}, \citenamefont {Aspelmeyer},\ and\ \citenamefont {Cirac}}]{Romero-Isart:2011sdw}%
  \BibitemOpen
  \bibfield  {author} {\bibinfo {author} {\bibfnamefont {O.}~\bibnamefont {Romero-Isart}}, \bibinfo {author} {\bibfnamefont {A.~C.}\ \bibnamefont {Pflanzer}}, \bibinfo {author} {\bibfnamefont {F.}~\bibnamefont {Blaser}}, \bibinfo {author} {\bibfnamefont {R.}~\bibnamefont {Kaltenbaek}}, \bibinfo {author} {\bibfnamefont {N.}~\bibnamefont {Kiesel}}, \bibinfo {author} {\bibfnamefont {M.}~\bibnamefont {Aspelmeyer}},\ and\ \bibinfo {author} {\bibfnamefont {J.~I.}\ \bibnamefont {Cirac}},\ }\bibfield  {title} {\bibinfo {title} {{Large Quantum Superpositions and Interference of Massive Nanometer-Sized Objects}},\ }\href {https://doi.org/10.1103/PhysRevLett.107.020405} {\bibfield  {journal} {\bibinfo  {journal} {Phys. Rev. Lett.}\ }\textbf {\bibinfo {volume} {107}},\ \bibinfo {pages} {020405} (\bibinfo {year} {2011})},\ \Eprint {https://arxiv.org/abs/1103.4081} {arXiv:1103.4081 [quant-ph]} \BibitemShut {NoStop}%
\bibitem [{\citenamefont {Schut}\ \emph {et~al.}(2025)\citenamefont {Schut}, \citenamefont {Andriolo}, \citenamefont {Toro{\v{s}}}, \citenamefont {Bose},\ and\ \citenamefont {Mazumdar}}]{Schut:2024lgp}%
  \BibitemOpen
  \bibfield  {author} {\bibinfo {author} {\bibfnamefont {M.}~\bibnamefont {Schut}}, \bibinfo {author} {\bibfnamefont {P.}~\bibnamefont {Andriolo}}, \bibinfo {author} {\bibfnamefont {M.}~\bibnamefont {Toro{\v{s}}}}, \bibinfo {author} {\bibfnamefont {S.}~\bibnamefont {Bose}},\ and\ \bibinfo {author} {\bibfnamefont {A.}~\bibnamefont {Mazumdar}},\ }\bibfield  {title} {\bibinfo {title} {{Expression for the decoherence rate due to air-molecule scattering in spatial qubits}},\ }\href {https://doi.org/10.1103/PhysRevA.111.042211} {\bibfield  {journal} {\bibinfo  {journal} {Phys. Rev. A}\ }\textbf {\bibinfo {volume} {111}},\ \bibinfo {pages} {042211} (\bibinfo {year} {2025})},\ \Eprint {https://arxiv.org/abs/2410.20910} {arXiv:2410.20910 [quant-ph]} \BibitemShut {NoStop}%
\bibitem [{\citenamefont {Toro\v{s}}\ \emph {et~al.}(2021)\citenamefont {Toro\v{s}}, \citenamefont {Van De~Kamp}, \citenamefont {Marshman}, \citenamefont {Kim}, \citenamefont {Mazumdar},\ and\ \citenamefont {Bose}}]{Toros:2020dbf}%
  \BibitemOpen
  \bibfield  {author} {\bibinfo {author} {\bibfnamefont {M.}~\bibnamefont {Toro\v{s}}}, \bibinfo {author} {\bibfnamefont {T.~W.}\ \bibnamefont {Van De~Kamp}}, \bibinfo {author} {\bibfnamefont {R.~J.}\ \bibnamefont {Marshman}}, \bibinfo {author} {\bibfnamefont {M.~S.}\ \bibnamefont {Kim}}, \bibinfo {author} {\bibfnamefont {A.}~\bibnamefont {Mazumdar}},\ and\ \bibinfo {author} {\bibfnamefont {S.}~\bibnamefont {Bose}},\ }\bibfield  {title} {\bibinfo {title} {{Relative acceleration noise mitigation for nanocrystal matter-wave interferometry: Applications to entangling masses via quantum gravity}},\ }\href {https://doi.org/10.1103/PhysRevResearch.3.023178} {\bibfield  {journal} {\bibinfo  {journal} {Phys. Rev. Res.}\ }\textbf {\bibinfo {volume} {3}},\ \bibinfo {pages} {023178} (\bibinfo {year} {2021})},\ \Eprint {https://arxiv.org/abs/2007.15029} {arXiv:2007.15029 [gr-qc]} \BibitemShut {NoStop}%
\bibitem [{\citenamefont {Schut}\ \emph {et~al.}(2024)\citenamefont {Schut}, \citenamefont {Bosma}, \citenamefont {Wu}, \citenamefont {Toro{\v{s}}}, \citenamefont {Bose},\ and\ \citenamefont {Mazumdar}}]{Schut:2023tce}%
  \BibitemOpen
  \bibfield  {author} {\bibinfo {author} {\bibfnamefont {M.}~\bibnamefont {Schut}}, \bibinfo {author} {\bibfnamefont {H.}~\bibnamefont {Bosma}}, \bibinfo {author} {\bibfnamefont {M.}~\bibnamefont {Wu}}, \bibinfo {author} {\bibfnamefont {M.}~\bibnamefont {Toro{\v{s}}}}, \bibinfo {author} {\bibfnamefont {S.}~\bibnamefont {Bose}},\ and\ \bibinfo {author} {\bibfnamefont {A.}~\bibnamefont {Mazumdar}},\ }\bibfield  {title} {\bibinfo {title} {{Dephasing due to electromagnetic interactions in spatial qubits}},\ }\href {https://doi.org/10.1103/PhysRevA.110.022412} {\bibfield  {journal} {\bibinfo  {journal} {Phys. Rev. A}\ }\textbf {\bibinfo {volume} {110}},\ \bibinfo {pages} {022412} (\bibinfo {year} {2024})},\ \Eprint {https://arxiv.org/abs/2312.05452} {arXiv:2312.05452 [quant-ph]} \BibitemShut {NoStop}%
\bibitem [{\citenamefont {Fragolino}\ \emph {et~al.}(2024)\citenamefont {Fragolino}, \citenamefont {Schut}, \citenamefont {Toro{\v{s}}}, \citenamefont {Bose},\ and\ \citenamefont {Mazumdar}}]{Fragolino:2023agd}%
  \BibitemOpen
  \bibfield  {author} {\bibinfo {author} {\bibfnamefont {P.}~\bibnamefont {Fragolino}}, \bibinfo {author} {\bibfnamefont {M.}~\bibnamefont {Schut}}, \bibinfo {author} {\bibfnamefont {M.}~\bibnamefont {Toro{\v{s}}}}, \bibinfo {author} {\bibfnamefont {S.}~\bibnamefont {Bose}},\ and\ \bibinfo {author} {\bibfnamefont {A.}~\bibnamefont {Mazumdar}},\ }\bibfield  {title} {\bibinfo {title} {{Decoherence of a matter-wave interferometer due to dipole-dipole interactions}},\ }\href {https://doi.org/10.1103/PhysRevA.109.033301} {\bibfield  {journal} {\bibinfo  {journal} {Phys. Rev. A}\ }\textbf {\bibinfo {volume} {109}},\ \bibinfo {pages} {033301} (\bibinfo {year} {2024})},\ \Eprint {https://arxiv.org/abs/2307.07001} {arXiv:2307.07001 [quant-ph]} \BibitemShut {NoStop}%
\bibitem [{\citenamefont {Mandel}\ and\ \citenamefont {Wolf}(1995)}]{mandel1995optical}%
  \BibitemOpen
  \bibfield  {author} {\bibinfo {author} {\bibfnamefont {L.}~\bibnamefont {Mandel}}\ and\ \bibinfo {author} {\bibfnamefont {E.}~\bibnamefont {Wolf}},\ }\href {https://doi.org/10.1017/CBO9781139644105} {\emph {\bibinfo {title} {Optical Coherence and Quantum Optics}}}\ (\bibinfo  {publisher} {Cambridge University Press},\ \bibinfo {address} {Cambridge, UK},\ \bibinfo {year} {1995})\BibitemShut {NoStop}%
\bibitem [{\citenamefont {Scully}\ and\ \citenamefont {Zubairy}(1997)}]{scully1997quantum}%
  \BibitemOpen
  \bibfield  {author} {\bibinfo {author} {\bibfnamefont {M.~O.}\ \bibnamefont {Scully}}\ and\ \bibinfo {author} {\bibfnamefont {M.~S.}\ \bibnamefont {Zubairy}},\ }\href {https://doi.org/10.1017/CBO9780511813993} {\emph {\bibinfo {title} {Quantum Optics}}}\ (\bibinfo  {publisher} {Cambridge University Press},\ \bibinfo {address} {Cambridge, UK},\ \bibinfo {year} {1997})\BibitemShut {NoStop}%
\bibitem [{\citenamefont {Schlosshauer}(2019)}]{schlosshauer2019quantum}%
  \BibitemOpen
  \bibfield  {author} {\bibinfo {author} {\bibfnamefont {M.}~\bibnamefont {Schlosshauer}},\ }\bibfield  {title} {\bibinfo {title} {Quantum decoherence},\ }\href {https://doi.org/10.1016/j.physrep.2019.10.001} {\bibfield  {journal} {\bibinfo  {journal} {Physics Reports}\ }\textbf {\bibinfo {volume} {831}},\ \bibinfo {pages} {1} (\bibinfo {year} {2019})}\BibitemShut {NoStop}%
\bibitem [{\citenamefont {Caves}(1981)}]{caves1981quantum}%
  \BibitemOpen
  \bibfield  {author} {\bibinfo {author} {\bibfnamefont {C.~M.}\ \bibnamefont {Caves}},\ }\bibfield  {title} {\bibinfo {title} {Quantum-mechanical noise in an interferometer},\ }\href {https://doi.org/10.1103/PhysRevD.23.1693} {\bibfield  {journal} {\bibinfo  {journal} {Physical Review D}\ }\textbf {\bibinfo {volume} {23}},\ \bibinfo {pages} {1693} (\bibinfo {year} {1981})}\BibitemShut {NoStop}%
\bibitem [{\citenamefont {Margalit}\ \emph {et~al.}(2021)\citenamefont {Margalit}, \citenamefont {Dobkowski}, \citenamefont {Zhou}, \citenamefont {Amit}, \citenamefont {Japha}, \citenamefont {Moukouri}, \citenamefont {Rohrlich}, \citenamefont {Mazumdar}, \citenamefont {Bose}, \citenamefont {Henkel} \emph {et~al.}}]{margalit2021realization}%
  \BibitemOpen
  \bibfield  {author} {\bibinfo {author} {\bibfnamefont {Y.}~\bibnamefont {Margalit}}, \bibinfo {author} {\bibfnamefont {O.}~\bibnamefont {Dobkowski}}, \bibinfo {author} {\bibfnamefont {Z.}~\bibnamefont {Zhou}}, \bibinfo {author} {\bibfnamefont {O.}~\bibnamefont {Amit}}, \bibinfo {author} {\bibfnamefont {Y.}~\bibnamefont {Japha}}, \bibinfo {author} {\bibfnamefont {S.}~\bibnamefont {Moukouri}}, \bibinfo {author} {\bibfnamefont {D.}~\bibnamefont {Rohrlich}}, \bibinfo {author} {\bibfnamefont {A.}~\bibnamefont {Mazumdar}}, \bibinfo {author} {\bibfnamefont {S.}~\bibnamefont {Bose}}, \bibinfo {author} {\bibfnamefont {C.}~\bibnamefont {Henkel}}, \emph {et~al.},\ }\bibfield  {title} {\bibinfo {title} {Realization of a complete stern-gerlach interferometer: Toward a test of quantum gravity},\ }\href {https://doi.org/10.1126/sciadv.abg2879} {\bibfield  {journal} {\bibinfo  {journal} {Science advances}\ }\textbf {\bibinfo {volume} {7}},\ \bibinfo {pages} {eabg2879} (\bibinfo {year} {2021})}\BibitemShut {NoStop}%
\bibitem [{\citenamefont {Machluf}\ \emph {et~al.}(2013)\citenamefont {Machluf}, \citenamefont {Japha},\ and\ \citenamefont {Folman}}]{Folman2013}%
  \BibitemOpen
  \bibfield  {author} {\bibinfo {author} {\bibfnamefont {S.}~\bibnamefont {Machluf}}, \bibinfo {author} {\bibfnamefont {Y.}~\bibnamefont {Japha}},\ and\ \bibinfo {author} {\bibfnamefont {R.}~\bibnamefont {Folman}},\ }\bibfield  {title} {\bibinfo {title} {Coherent stern--gerlach momentum splitting on an atom chip},\ }\href {http://dx.doi.org/10.1038/ncomms3424} {\bibfield  {journal} {\bibinfo  {journal} {Nature Communications}\ }\textbf {\bibinfo {volume} {4}},\ \bibinfo {pages} {2424} (\bibinfo {year} {2013})}\BibitemShut {NoStop}%
\bibitem [{\citenamefont {Margalit}\ \emph {et~al.}(2018)\citenamefont {Margalit}, \citenamefont {Zhou}, \citenamefont {Dobkowski}, \citenamefont {Japha}, \citenamefont {Rohrlich}, \citenamefont {Moukouri},\ and\ \citenamefont {Folman}}]{Folman2018}%
  \BibitemOpen
  \bibfield  {author} {\bibinfo {author} {\bibfnamefont {Y.}~\bibnamefont {Margalit}}, \bibinfo {author} {\bibfnamefont {Z.}~\bibnamefont {Zhou}}, \bibinfo {author} {\bibfnamefont {O.}~\bibnamefont {Dobkowski}}, \bibinfo {author} {\bibfnamefont {Y.}~\bibnamefont {Japha}}, \bibinfo {author} {\bibfnamefont {D.}~\bibnamefont {Rohrlich}}, \bibinfo {author} {\bibfnamefont {S.}~\bibnamefont {Moukouri}},\ and\ \bibinfo {author} {\bibfnamefont {R.}~\bibnamefont {Folman}},\ }\bibfield  {title} {\bibinfo {title} {Realization of a complete stern-gerlach interferometer},\ }\href@noop {} {\bibfield  {journal} {\bibinfo  {journal} {arXiv preprint arXiv:1801.02708}\ } (\bibinfo {year} {2018})}\BibitemShut {NoStop}%
\bibitem [{\citenamefont {Amit}\ \emph {et~al.}(2019)\citenamefont {Amit}, \citenamefont {Margalit}, \citenamefont {Dobkowski}, \citenamefont {Zhou}, \citenamefont {Japha}, \citenamefont {Zimmermann}, \citenamefont {Efremov}, \citenamefont {Narducci}, \citenamefont {Rasel}, \citenamefont {Schleich},\ and\ \citenamefont {Folman}}]{folman2019}%
  \BibitemOpen
  \bibfield  {author} {\bibinfo {author} {\bibfnamefont {O.}~\bibnamefont {Amit}}, \bibinfo {author} {\bibfnamefont {Y.}~\bibnamefont {Margalit}}, \bibinfo {author} {\bibfnamefont {O.}~\bibnamefont {Dobkowski}}, \bibinfo {author} {\bibfnamefont {Z.}~\bibnamefont {Zhou}}, \bibinfo {author} {\bibfnamefont {Y.}~\bibnamefont {Japha}}, \bibinfo {author} {\bibfnamefont {M.}~\bibnamefont {Zimmermann}}, \bibinfo {author} {\bibfnamefont {M.~A.}\ \bibnamefont {Efremov}}, \bibinfo {author} {\bibfnamefont {F.~A.}\ \bibnamefont {Narducci}}, \bibinfo {author} {\bibfnamefont {E.~M.}\ \bibnamefont {Rasel}}, \bibinfo {author} {\bibfnamefont {W.~P.}\ \bibnamefont {Schleich}},\ and\ \bibinfo {author} {\bibfnamefont {R.}~\bibnamefont {Folman}},\ }\bibfield  {title} {\bibinfo {title} {${T}^{3}$ stern-gerlach matter-wave interferometer},\ }\href {https://doi.org/10.1103/PhysRevLett.123.083601} {\bibfield  {journal} {\bibinfo  {journal} {Phys. Rev. Lett.}\ }\textbf {\bibinfo {volume} {123}},\ \bibinfo {pages} {083601} (\bibinfo
  {year} {2019})}\BibitemShut {NoStop}%
\bibitem [{\citenamefont {Gram}(1883)}]{Gram1883}%
  \BibitemOpen
  \bibfield  {author} {\bibinfo {author} {\bibfnamefont {J.~P.}\ \bibnamefont {Gram}},\ }\bibfield  {title} {\bibinfo {title} {Ueber die entwickelung reeller functionen in reihen mittelst der methode der kleinsten quadrate},\ }\href@noop {} {\bibfield  {journal} {\bibinfo  {journal} {Journal f{\"u}r die reine und angewandte Mathematik}\ }\textbf {\bibinfo {volume} {94}},\ \bibinfo {pages} {41} (\bibinfo {year} {1883})}\BibitemShut {NoStop}%
\bibitem [{\citenamefont {Schmidt}(1907)}]{Schmidt1907}%
  \BibitemOpen
  \bibfield  {author} {\bibinfo {author} {\bibfnamefont {E.}~\bibnamefont {Schmidt}},\ }\bibfield  {title} {\bibinfo {title} {Zur theorie der linearen und nichtlinearen integralgleichungen. i. teil: Entwicklung willk{\"u}rlicher functionen nach systemen vorgeschriebener},\ }\href@noop {} {\bibfield  {journal} {\bibinfo  {journal} {Mathematische Annalen}\ }\textbf {\bibinfo {volume} {63}},\ \bibinfo {pages} {433} (\bibinfo {year} {1907})}\BibitemShut {NoStop}%
\bibitem [{\citenamefont {Leon}\ \emph {et~al.}(2013)\citenamefont {Leon}, \citenamefont {Bj{\"o}rck},\ and\ \citenamefont {Gander}}]{Leon2013}%
  \BibitemOpen
  \bibfield  {author} {\bibinfo {author} {\bibfnamefont {S.~J.}\ \bibnamefont {Leon}}, \bibinfo {author} {\bibfnamefont {{\AA}.}~\bibnamefont {Bj{\"o}rck}},\ and\ \bibinfo {author} {\bibfnamefont {W.}~\bibnamefont {Gander}},\ }\bibfield  {title} {\bibinfo {title} {Gram-schmidt orthogonalization: 100 years and more},\ }\href {https://doi.org/10.1002/nla.1839} {\bibfield  {journal} {\bibinfo  {journal} {Numerical Linear Algebra with Applications}\ }\textbf {\bibinfo {volume} {20}},\ \bibinfo {pages} {492} (\bibinfo {year} {2013})}\BibitemShut {NoStop}%
\bibitem [{\citenamefont {Wolf}(1959)}]{Wolf1959}%
  \BibitemOpen
  \bibfield  {author} {\bibinfo {author} {\bibfnamefont {E.}~\bibnamefont {Wolf}},\ }\bibfield  {title} {\bibinfo {title} {Coherence properties of optical fields},\ }\href@noop {} {\bibfield  {journal} {\bibinfo  {journal} {Nuovo Cimento}\ }\textbf {\bibinfo {volume} {13}},\ \bibinfo {pages} {1165} (\bibinfo {year} {1959})}\BibitemShut {NoStop}%
\bibitem [{\citenamefont {Jauch}\ and\ \citenamefont {Rohrlich}(1980)}]{Jauch1980}%
  \BibitemOpen
  \bibfield  {author} {\bibinfo {author} {\bibfnamefont {J.~M.}\ \bibnamefont {Jauch}}\ and\ \bibinfo {author} {\bibfnamefont {F.}~\bibnamefont {Rohrlich}},\ }\href@noop {} {\emph {\bibinfo {title} {The Theory of Photons and Electrons}}}\ (\bibinfo  {publisher} {Springer-Verlag},\ \bibinfo {address} {Berlin},\ \bibinfo {year} {1980})\ \bibinfo {note} {sec. 2.8}\BibitemShut {NoStop}%
\bibitem [{\citenamefont {Born}\ and\ \citenamefont {Wolf}(1999)}]{BornWolf1999}%
  \BibitemOpen
  \bibfield  {author} {\bibinfo {author} {\bibfnamefont {M.}~\bibnamefont {Born}}\ and\ \bibinfo {author} {\bibfnamefont {E.}~\bibnamefont {Wolf}},\ }\href@noop {} {\emph {\bibinfo {title} {Principles of Optics: Electromagnetic Theory of Propagation, Interference and Diffraction of Light}}},\ \bibinfo {edition} {7th}\ ed.\ (\bibinfo  {publisher} {Cambridge University Press},\ \bibinfo {address} {Cambridge},\ \bibinfo {year} {1999})\ \bibinfo {note} {sec. 1.4}\BibitemShut {NoStop}%
\bibitem [{\citenamefont {Abouraddy}\ \emph {et~al.}(2002)\citenamefont {Abouraddy}, \citenamefont {Sergienko}, \citenamefont {Saleh},\ and\ \citenamefont {Teich}}]{Abouraddy2002}%
  \BibitemOpen
  \bibfield  {author} {\bibinfo {author} {\bibfnamefont {A.~F.}\ \bibnamefont {Abouraddy}}, \bibinfo {author} {\bibfnamefont {A.~V.}\ \bibnamefont {Sergienko}}, \bibinfo {author} {\bibfnamefont {B.~E.~A.}\ \bibnamefont {Saleh}},\ and\ \bibinfo {author} {\bibfnamefont {M.~C.}\ \bibnamefont {Teich}},\ }\bibfield  {title} {\bibinfo {title} {Quantum-entanglement-based binary optical communications},\ }\href@noop {} {\bibfield  {journal} {\bibinfo  {journal} {Optics Communications}\ }\textbf {\bibinfo {volume} {201}},\ \bibinfo {pages} {93} (\bibinfo {year} {2002})}\BibitemShut {NoStop}%
\bibitem [{\citenamefont {Dutta}\ \emph {et~al.}(2021)\citenamefont {Dutta}, \citenamefont {Ghosh}, \citenamefont {Kim},\ and\ \citenamefont {Sengupta}}]{dutta2021robust}%
  \BibitemOpen
  \bibfield  {author} {\bibinfo {author} {\bibfnamefont {A.}~\bibnamefont {Dutta}}, \bibinfo {author} {\bibfnamefont {S.}~\bibnamefont {Ghosh}}, \bibinfo {author} {\bibfnamefont {J.}~\bibnamefont {Kim}},\ and\ \bibinfo {author} {\bibfnamefont {R.}~\bibnamefont {Sengupta}},\ }\bibfield  {title} {\bibinfo {title} {Robust entanglement detection in arbitrary two-mode {G}aussian state: {A} {S}tokes-like operator-based approach},\ }\href {https://arxiv.org/abs/2103.12987} {\bibfield  {journal} {\bibinfo  {journal} {arXiv preprint arXiv:2103.12987}\ } (\bibinfo {year} {2021})}\BibitemShut {NoStop}%
\bibitem [{\citenamefont {Leonhardt}(1997)}]{leonhardt_measuring_1997}%
  \BibitemOpen
  \bibfield  {author} {\bibinfo {author} {\bibfnamefont {U.}~\bibnamefont {Leonhardt}},\ }\href@noop {} {\emph {\bibinfo {title} {Measuring the Quantum State of Light}}},\ \bibinfo {series} {Cambridge Studies in Modern Optics}, Vol.~\bibinfo {volume} {22}\ (\bibinfo  {publisher} {Cambridge University Press},\ \bibinfo {address} {Cambridge},\ \bibinfo {year} {1997})\BibitemShut {NoStop}%
\bibitem [{\citenamefont {Welsch}\ \emph {et~al.}(1999)\citenamefont {Welsch}, \citenamefont {Vogel},\ and\ \citenamefont {Opatrn{\'y}}}]{welsch1999homodyne}%
  \BibitemOpen
  \bibfield  {author} {\bibinfo {author} {\bibfnamefont {D.-G.}\ \bibnamefont {Welsch}}, \bibinfo {author} {\bibfnamefont {W.}~\bibnamefont {Vogel}},\ and\ \bibinfo {author} {\bibfnamefont {T.}~\bibnamefont {Opatrn{\'y}}},\ }\bibfield  {title} {\bibinfo {title} {Homodyne detection and quantum state reconstruction},\ }in\ \href {https://doi.org/10.1016/S0079-6638(08)70389-5} {\emph {\bibinfo {booktitle} {Progress in Optics}}},\ Vol.~\bibinfo {volume} {39},\ \bibinfo {editor} {edited by\ \bibinfo {editor} {\bibfnamefont {E.}~\bibnamefont {Wolf}}}\ (\bibinfo  {publisher} {Elsevier},\ \bibinfo {address} {Amsterdam},\ \bibinfo {year} {1999})\ pp.\ \bibinfo {pages} {63--211}\BibitemShut {NoStop}%
\bibitem [{\citenamefont {Walls}\ and\ \citenamefont {Milburn}(2008)}]{WallsMilburn2008}%
  \BibitemOpen
  \bibfield  {author} {\bibinfo {author} {\bibfnamefont {D.~F.}\ \bibnamefont {Walls}}\ and\ \bibinfo {author} {\bibfnamefont {G.~J.}\ \bibnamefont {Milburn}},\ }\href {https://doi.org/10.1007/978-3-540-28574-8} {\emph {\bibinfo {title} {Quantum Optics}}},\ \bibinfo {edition} {2nd}\ ed.\ (\bibinfo  {publisher} {Springer-Verlag},\ \bibinfo {address} {Berlin, Heidelberg},\ \bibinfo {year} {2008})\BibitemShut {NoStop}%
\bibitem [{\citenamefont {Werner}(1989)}]{Werner1989}%
  \BibitemOpen
  \bibfield  {author} {\bibinfo {author} {\bibfnamefont {R.~F.}\ \bibnamefont {Werner}},\ }\bibfield  {title} {\bibinfo {title} {Quantum states with einstein-podolsky-rosen correlations admitting a hidden-variable model},\ }\href {https://doi.org/10.1103/PhysRevA.40.4277} {\bibfield  {journal} {\bibinfo  {journal} {Physical Review A}\ }\textbf {\bibinfo {volume} {40}},\ \bibinfo {pages} {4277} (\bibinfo {year} {1989})}\BibitemShut {NoStop}%
\end{thebibliography}%
\end{document}